\documentclass[11pt, a4paper]{article}

\usepackage{epsfig}
\usepackage{authblk}
\usepackage{amsmath}
\usepackage{siunitx}
\usepackage{natbib}
\usepackage{authblk}
\usepackage{tabularx}

\title{Quantification of plastic deformation in anorthite using micropillar compression}

\author[1]{Sarah Incel}
\author[2]{Chunhua Tian}
\author[1]{Anja Schreiber}
\author[1]{Vladimir Roddatis}
\author[2]{Sandra Korte-Kerzel}
\affil[1]{GFZ Helmholtz Centre for Geosciences, Telegrafenberg, 14473 Potsdam, Germany}
\affil[2]{RWTH Aachen University, Institute for Physical Metallurgy and Materials Physics, Kopernikusstraße 14, 52074 Aachen, Germany}

\date{}

\begin{document}
  \maketitle

\begin{abstract}
Plagioclase is the most abundant mineral group of the crust and will therefore either control or strongly influence crustal rheology. Yet, we still know little about crystal plasticity in plagioclase as brittle deformation dominates under most laboratory conditions. To overcome past challenges in quantifying crystal plastic deformation in plagioclase, we used micropillar compression to quantify the strength of specific slip systems as well as the stresses necessary for mechanical twinning in anorthite. Pillars with diameters of around 1 µm were milled from three differently oriented cuts of the same anorthite single crystals using focused-ion beam. Uniaxial deformation of 36 pillars in total was then conducted inside a scanning-electron microscope at either 25, 300, or 800 $^{\circ}$C. Most pillars were deformed with a constant displacement rate that corresponds to a strain rate of around 10$^{-3}$ s$^{-1}$ on the pillars. In addition, at 800 $^{\circ}$C, pillars were deformed with strain rates of 10$^{-2}$ or 10$^{-1}$ s$^{-1}$, respectively, to test the sensitivity of the deformation-controlling mechanism to changes in strain rate. Pillars deform in a plastic manner under all temperatures and strain rates tested and yield at different stresses depending on their orientation. Variations in strain rate of one to two orders of magnitude seem to have no influence on pillar strength. The critical resolved shear stress (CRSS) for mechanical (pericline) twinning lies at around 300 MPa and is, as expected from previous studies on other materials, insensitive to changes in temperature or strain rate. Pillar deformation in crystals oriented to activate slip exhibit no changes in yield stress with changes in strain rate over two orders of magnitude, but a strong decrease in yield stress with increasing temperature. Among the deformation mechanisms active during pillar deformation, twinning is the easiest under the applied conditions, followed by slip on \{011\}$\langle$100$\rangle$ regardless of temperature. Slip on \{011\}$\langle0\bar{1}1\rangle$ or $\{0\bar{1}1\}\langle011\rangle$ is consistently the hardest mechanism to activate in our anorthite pillars.
\end{abstract}

\section{Introduction}

The crust of the Earth, and of other terrestrial bodies such as the Earth's moon \citep[e.g.,][]{Wieczorek2001}, Mars \citep[e.g.,][]{McSween2009}, Venus \citep[e.g., see Fig. 3 in][]{Collinet2025}, and Mercury \cite[e.g.,][]{Namur2017} are all rich in plagioclase. Earth's crust is made up by roughly 40\% of plagioclase \citep[][]{Ronov1969}, which forms a continuous solid solution between the Na-rich endmember albite (Ab) and the Ca-rich endmember anorthite (An). Despite its abundance in the continental crust, especially at mid- to lower depths, around 15 to 80 km, and the resulting importance for crustal rheology, there is a lack in knowledge in how plagioclase deforms under those elevated temperatures and pressures. As we expect that plastic deformation mechanisms become more important in the deformation of rocks with increasing pressure, and especially with increasing temperature \citep[e.g.,][]{Paterson1958}, the way plagioclase deforms in the plastic regime is likely to control the strength of the mid to lower continental crust. The pioneering work by \citet{Rybacki2000} lead to the establishment of a flow law for 'wet' and 'dry' anorthite. Despite this indisputable achievement, the use of flow laws relying on the relation between strain rate and material flow under constant stress, remains probably restricted to settings in which we assume constant load conditions. Yet, in many geological settings we observe that the continental lithosphere is exposed to continuous changes in load, e.g., due to melting ice sheets (glacial isostatic adjustment) or after large earthquakes (postseismic relaxation). Hence, applying a constitutive law derived from a mechanical steady-state (where d$\sigma$/d\emph{t} = 0) to geological settings which are better described by either constantly or transiently changing loads bears risks as viscosities may be over- or underestimated. 

Many models assume that during the first stage of deformation, strain is accommodated by dislocation motion on the easy slip system whereas the hard slip system becomes rate controlling at larger strains \citep[e.g.,][]{Karato2008, Masuti2016, muto2019, Masuti2023}. The concept of such a stress transfer in the isostrain endmember model, on which these studies are based, was recently challenged by \citet{Wallis2025}, who argue that isostrain models cannot capture strain hardening observed in experimental studies \citep{Hansen2019, Hansen2021}. The authors assume that hardening mainly results from the elastic interaction between dislocations (backstress), and by taking this process into account leads to an excellent agreement between laboratory data and measurements of postseismic creep \citep{Breithaupt2023}. Nevertheless, regardless of the underlying model, all studies demonstrate that by knowing the strength of individual slip systems, one can simulate the evolution of viscosity with strain. Such simulations are extremely valuable, for large strains are generally challenging to achieve through laboratory experiments. Furthermore, using viscoplastic self-consistent modelling, the knowledge about the relative strengths between slip systems was used to simulate continental rifting \citep{Tommasi2001}, the formation of proto-plate boundaries \citep{Tommasi2009}, or seismic anisotropies in the upper mantle \citep[e.g.,][]{Mainprice2005}. 

To quantify the stress necessary to activate a specific slip system, that is the motion of dislocations in a crystallographic plane along a certain direction, experiments on oriented single crystals are required as they allow us to control the stresses resolved on specific planes in distinct directions. As plagioclase shows perfect and good cleavage along (001) and (010), respectively, cleavage cracking appears dominant over a wide range of pressures and temperatures accessible with standard rock deformation techniques \citep{Incel2024, Tullis1992}. Hence, the only two previous attempts, known to the authors, at compressing plagioclase single crystals under high confinement (1 GPa) and elevated temperature (700 to 900 $^{\circ}$C) resulted in a mixture of deformation mechanisms. \citet{Stunitz2003} report the occurrence of cracking in their single crystals, even when samples were exposed to strain rates of 10$^{-7}$ s$^{-1}$, considered very low for experimental deformation studies. \citet{Borg1969} found the simultaneous occurrence of twinning and dislocation glide, although their tests were conducted at significantly higher strain rates, i.e., 10$^{-5}$ to 10$^{-4}$ s$^{-1}$. Several experimental studies on polycrystalline plagioclase aggregates further demonstrate that brittle deformation dominantes under most pressure, temperature, and strain rate conditions commonly selected for deformation tests \citep[e.g.,][and references therein]{Tullis1992, Incel2019, Incel2024, Baisset2024}. Since the experimental methods did not evolve over the past decades, quantifying plastic deformation mechanisms in plagioclase using standard rock deformation apparatus remains challenging or even impossible. 

Besides dislocation motion, deformation by twinning can be a material's response to an external stress as well. Mechanical or deformation twinning is the process of simultaneous shear of a volume within a crystal and only occurs along specific crystallographic directions. Plagioclase, like many low-symmetry phases, seems to easily deform by twinning as mechanical twinning is often the first plastic deformation mechanism active in plagioclase-rich rocks. \citet{Tullis1992} documented that mechanical twinning in plagioclase starts in their Bushveld anorthosites at a temperature of around 500 $^{\circ}$C whereas the first evidence for crystal plasticity was found in samples deformed at 900 $^{\circ}$C. In experimentally deformed plagioclase single crystals, \citet{Stunitz2003} and \citet{Marshall1977} found that most twins originated from microcracks or (sub-)grain boundaries, a relation commonly observed in naturally deformed rocks as well \citep[see Fig. 2 in][]{Olsen1985}. Due to the close spatial relation between cracks or grain boundaries and twins, it appears that twin nucleation is facilitated by the presence of defects either as they are associated with elevated stresses or may aid in the required  structural re-arrangement between twin and parent \citep[p. 45 in][]{Christian1995}. Mechanical twinning in plagioclase either obeys the albite or pericline twin law. The former describes a reflection about the (010) plane whereas the latter is produced by a rotation of 180$^{\circ}$ around the [010] direction \citep[e.g.,][]{Nespolo2025}. A recent publication describes the process of mechanical twinning in plagioclase in much detail \citep{Xie2019}. The ease at which plagioclase can deform by twinning is discussed to change as a function of chemical composition within the plagioclase solid solution and thus with Al/Si order \citep{Brown1986}. 
Due to the insensitivity of deformation twinning to changes in temperature and strain rate, calcite deformation twins have been extensively used to estimate paleo stresses in natural rocks \citep[e.g.,][and references therein]{Rowe1990, Lacombe2001, Lacombe2007a, Behr2014a, Lacombe2021a, Rutter2022}, which was made possible thanks to the seminal work on calcite plasticity by \citet{DeBresser1993, DeBresser1997} and by \citet{Turner1954}. 

We deformed single crystals of almost pure anorthite (An$_{96}$) with the goal to quantify the stress necessary to activate specific slip systems as well as deformation twinning. To overcome past challenges in quantifying plastic deformation in hard and brittle materials, such as ceramics \citep[e.g.,][and references therein]{Korte-Kerzel2017}, we used micropillar compression, a method commonly used in the materials science community. Previous results on oxides such as spinel \citep{Korte2009} as well as on calcite \citep{Sly2020} and olivine \citep{Kranjc2020} exhibit that micropillar compression is suitable to investigate and to determine dislocation-mediated deformation in hard and brittle Earth materials as well. Here, we used micropillar compression to quantify the stresses necessary to activate deformation twinning, and slip along \{011\}$\langle100\rangle$ and \{011\}$\langle1\bar{1}1\rangle$ or \{0$\bar{1}$1\}$\langle$111$\rangle$, respectively. We further found that pillars orientation has a major influence on the overall deformation behaviour of the pillars as well as on their ductility.

\section{Materials \& Methods}
In the present study we used a natural anorthite crystal from Yoichi, Hokkaido, Japan, as starting material. This crystal revealed clear facets, which informed sample preparation to obtain three cuts in total. Two cuts were oriented in a way to activate slip (orientation 1 and 3 hereafter), i.e., surface normal is perpendicular to either \{110\} or \{001\}, respectively, and one was oriented favourably for twinning (orientation 2), i.e., the normal to the surface is perpendicular to \{$\bar{1}$10\}. All crystals were polished using SiC paper (2500 grid), diamond paste (1 µm), and finally in suspension (0.25 µm) on a vibropolisher for at least 2 h. Subsequently, the crystals were carbon coated for inspection under the electron microprobe (JEOL Hyperprobe JXA-8530F). Using energy- as well as angle-dispersive X-ray spectroscopy, we found that besides a few pores surrounded by a K-rich material, the starting material is chemically homogeneous. The exact formula of the staring material is Ca$_{0.93}$Na$_{0.04}$Al$_{1.95}$Si$_{2}$O$_{8}$ with traces of Mg, Fe$^{2+}$ and K, resulting in an anorthite fraction of An$_{96}$ (i.e., An/An+Ab). Thus, the starting material is almost pure anorthite. 

From the three differently oriented samples, we then produced pillars by using focused-ion beam (FIB) using a FEI - Helios G4 UC - Dual Beam scanning electron microscope (SEM) as well as a Helios NanoLab 600i DualBeam. To obtain free-standing cylindrical pillars, we removed material stepwise by using Ga ions. First, we milled a circular area with an outer diameter of around 15 or 25 µm, an inner diameter of 10 µm, and a depth of around 2 µm using a current of 9.3 nA. During the second and third milling step we reduced the pillar diameter down to 5 and then 2 µm, and the beam current to 2.5 and 0.79 nA. Finally, we removed the last material to obtain pillars of around 1 µm with a beam current of 80 pA and to keep beam damage to a minimum. The acceleration voltage was kept constant at 30 keV. These milling steps were generally sufficient to produce a pillar with the target dimension of 1 µm diameter and 2.5 µm height in the centre of the circular pits. In some cases, we needed an additional milling step, because of left-over material on some pillar flanks, resulting in pillars becoming longer than the desired 2.5 µm, and eventually in the pillars revealing a range in height from around 2.5 to 4 µm. Since the diameter was with around 1$\pm$0.1 µm kept constant, the aspect ratio of the pillars tested varied. Pillars revealed a taper of 2 to 3$^{\circ}$. Yet, neither taper nor changes in aspect ratio did have any noticeable influence on the mechanical data, e.g., on buckling under load or on strength. 

Micropillar compression tests were conducted using two different mechanical testing stages, a FT-NMT04-XYZ in-situ nanoindenter from FemtoTools and an Alemnis Standard Assembly equipped with a high-temperture module. The use of two different experimental setups was motivated by the latter setup allowing for deformation at high temperatures of up to 1000 $^{\circ}$C. Both mechanical testing stages are mounted in an SEM enabling us to record load and displacement while observing pillar deformation. We deformed pillars at three different temperatures, i.e., 25, 300, or 800 $^{\circ}$C by advancing a diamond flat punch indenter (5 µm diameter) with a controlled displacement rate. Most tests were conducted at a displacement rate that corresponds to a strain rate of 10$^{-3}$ s$^{-1}$ on the pillar. To test strain-rate sensitivity, we performed additional tests with strain rates of 10$^{-2}$  for samples with orientations 1 and 2 and with 10$^{-2}$ and 10$^{-1}$ s$^{-1}$ for the sample with orientation 3. The strain rate sensitivity tests were all conducted at 800 $^{\circ}$C. Displacement is measured using a piezotransducer with a resolution of around 1 nm. Load is measured using a strain gage load cell and its resolution is around 1 µN. The error on temperature measurements is negligible below, and around 10 $^{\circ}$C above 800 $^{\circ}$C. The sampling rate for all measurements was 20 Hz. 

In total, we deformed 36 anorthite micropillars. Prior to the tests, we measured the initial pillar length \emph{l}$_{0}$ and pillar diameter \emph{d}$_{0}$, which both have an uncertainty of around 10\%. We used the measured load to obtain the evolution of true stress, i.e., $\mathrm{\sigma_{true} = \sigma_{eng} (1 - \epsilon_{eng})}$, with true strain, i.e., $\mathrm{\epsilon_{true} = -ln(1 - \epsilon_{eng})}$, taking into account the changing pillar diameter while we uniaxially compressed the pillars during deformation at constant displacement rate. In most tests, we observe a distinct load drop after a few percent strain. This drop in load marks the yield stress. In a few high-temperature tests, in which we observe continuous flow, we used the deviation from linearity, marking the transition from elastic to plastic behaviour as yield point. 

After deformation, we selected several pillars to investigate their respective nanostructures using an transmission-electron microscope (TEM; Thermo Fisher Scientific Themis Z). First, FIB foils were cut to reveal the pillars' cross-sections. Subsequently, the foils were investigated at the TEM. We used selected area electron diffraction (SAED) pattern, in conjunction with the softwares imageJ, Vesta, SingleCrystal, and CrystalMaker to obtain the initial pillar orientation by indexing the SAED pattern, and to extract any information on orientation relationships between undeformed and deformed areas in the pillars, e.g., between twin and parent. 

\section{Results}
\label{}
In the following, we will present the mechanical data of 36 micropillar compression tests and exhibit the nanostructures of nine pillars after deformation. For the latter, we selected three pillars per set of orientation, and each of the three pillars was deformed at a different temperature, i.e., at 25, 300 or 800 $^{\circ}$C. Throughout the manuscript mechanical data and nanostructural analyses are colour-coded to indicate the temperature under which the respective pillars were deformed. Blue marks deformation at 25 $^{\circ}$C, whereas yellow and red reflect deformation at 300 or 800 $^{\circ}$C, respectively. Please note that, if not stated otherwise, the use of the words strain and stress refer to true strain and true stress, respectively. The experimental conditions each pillar experienced as well as their initial orientation, yield stress, and total strain are listed in Table \ref{ExpCond}.

\begin{table}
\caption{Experimental conditions and mechanical results of the 36 micropillar compression tests. Loading direction was perpendicular to the listed crystallographic planes. SD= standard deviation}
\label{ExpCond}
\begin{tabularx}{\textwidth}{X X X X X X X}
Run  & Loading direction& \emph{T} [K] & Strain rate  [s$^{-1}$] & Yield stress$\pm$SD [GPa] & Total strain [\%]\\
\hline
 O1P1$^{\circ}$  & \{110\} & 298 & 10$^{-3}$ & 4.2$\pm$0.3 & 14.14 \\
 O1P2  & \{110\} & 298 & 10$^{-3}$ & 5.1$\pm$0.3 & 17.22 \\
 O1P3$^\heartsuit$  & \{110\} & 298 & 10$^{-3}$ & 4.7$\pm$0.3 & 17.67 \\
 O1P4$^\heartsuit$  & \{110\} & 298 & 10$^{-3}$ & 5.1$\pm$0.3 & 9.35 \\
 O1P5  & \{110\} & 298 & 10$^{-3}$ & 4.8$\pm$0.3 & 10.58 \\
 O1P6  & \{110\} & 298 & 10$^{-3}$ & 4.6$\pm$0.3 & 8.43 \\
 O1P7  & \{110\} & 573 & 10$^{-3}$ & 3.7$\pm$0.1 & 24.35 \\
 O1P8$^{\circ}$  & \{110\} & 573 & 10$^{-3}$ & 3.4$\pm$0.1 & 12.12 \\
 O1P9  & \{110\} & 1073 & 10$^{-3}$ & 2.0$\pm$0.1 & 23.77 \\
 O1P10$^{\circ}$  & \{110\} & 1073 & 10$^{-3}$ & 2.2$\pm$0.1 & 13.78 \\
 O1P11  & \{110\} & 1073 & 10$^{-3}$ & 2.2$\pm$0.1 & 13.01\\
 O1P12  & \{110\} & 1073 & 10$^{-3}$ & 2.0$\pm$0.1 & 8.24 \\
 O1P13  & \{110\} & 1073 & 10$^{-2}$ & 2.1 & 12.42\\
 \hline
 O2P1  & \{$\bar{1}$10\} & 298 & 10$^{-3}$ & 1.5$\pm$0.3& 17.70 \\ 
 O2P2  & \{$\bar{1}$10\} &  298 & 10$^{-3}$ & 1.1$\pm$0.3 & 29.43 \\ 
 O2P3  & \{$\bar{1}$10\} & 298 & 10$^{-3}$ & 1.1$\pm$0.3 & 30.13 \\ 
 O2P4  & \{$\bar{1}$10\} & 298 & 10$^{-3}$ & 0.7$\pm$0.3 & 12.78 \\ 
 O2P5*  & \{$\bar{1}$10\} & 298 & 10$^{-3}$ & 0.3$\pm$0.3 & 5.65 \\ 
 O2P6$^{\circ}$  & \{$\bar{1}$10\} & 298 & 10$^{-3}$ & 1.5$\pm$0.3 & 6.72 \\ 
 O2P7  & \{$\bar{1}$10\} & 573 & 10$^{-3}$ & 1.3$\pm$0.3 & 26.25 \\ 
 O2P8  & \{$\bar{1}$10\} & 573 & 10$^{-3}$ & 1.2$\pm$0.3 & 9.68 \\
 O2P9$^{\circ}$  & \{$\bar{1}$10\} & 573 & 10$^{-3}$ & 1.8$\pm$0.3 & 21.83\\
 O2P10  & \{$\bar{1}$10\} & 1073 & 10$^{-3}$ & 1.1$\pm$0.2& 11.58\\
 O2P11$^{\circ}$  & \{$\bar{1}$10\} & 1073 & 10$^{-3}$ & 0.8$\pm$0.2 & 11.57 \\
 O2P12  & \{$\bar{1}$10\} & 1073 & 10$^{-2}$ & 1.1 & 18.03 \\
 \hline
 O3P1   & \{001\} & 298 & 10$^{-3}$ & 5.9$\pm$0.4 & 9.20 \\
 O3P2  & \{001\} & 298 & 10$^{-3}$ & 5.2$\pm$0.4 & 15.21 \\
 O3P3$^{\circ}$  & \{001\} & 298 & 10$^{-3}$ & 5.1$\pm$0.4 & 28.92 \\
 O3P4  & \{001\} & 298 & 10$^{-3}$ & 6.2$\pm$0.4 & 6.88 \\
 O3P5  & \{001\} & 573 & 10$^{-3}$ & 4.1$\pm$0.5 & 22.64 \\
 O3P6$^{\circ}$  & \{001\} & 573 & 10$^{-3}$ & 5.1$\pm$0.5 & 8.97 \\
 O3P7$^{\circ}$  & \{001\} & 1073 & 10$^{-3}$ & 3.6$\pm$0.3 & 7.64 \\
 O3P8  & \{001\} & 1073 & 10$^{-3}$ & 4.2$\pm$0.3 & 9.54 \\
 O3P9  & \{001\} & 1073 & 10$^{-3}$ & 3.6$\pm$0.3 & 4.18 \\
 O3P10  & \{001\} & 1073 & 10$^{-2}$ & 7.5 & 9.81 \\
 O3P11  & \{001\} & 1073 & 10$^{-1}$ & 4.4 & 10.34\\
   \hline
\end{tabularx}
 \footnotesize{$^{\circ}$Corresponding nanostructures are exhibited in Figs. \ref{TEM1}; \ref{TEM2}; \ref{TEM3}. *This test shows no stress drop associated with twin nucleation, and therefore a substantially different deformation behaviour to all other pillars of orientation 2. We thus did not include the associated yield stress in the calculation of the SD. $^\heartsuit$Tests conducted with electron beam off.}
 \end{table}
 
\subsection{Mechanical data}
\begin{figure}[h] 
   \centering
   \includegraphics[width=13cm]{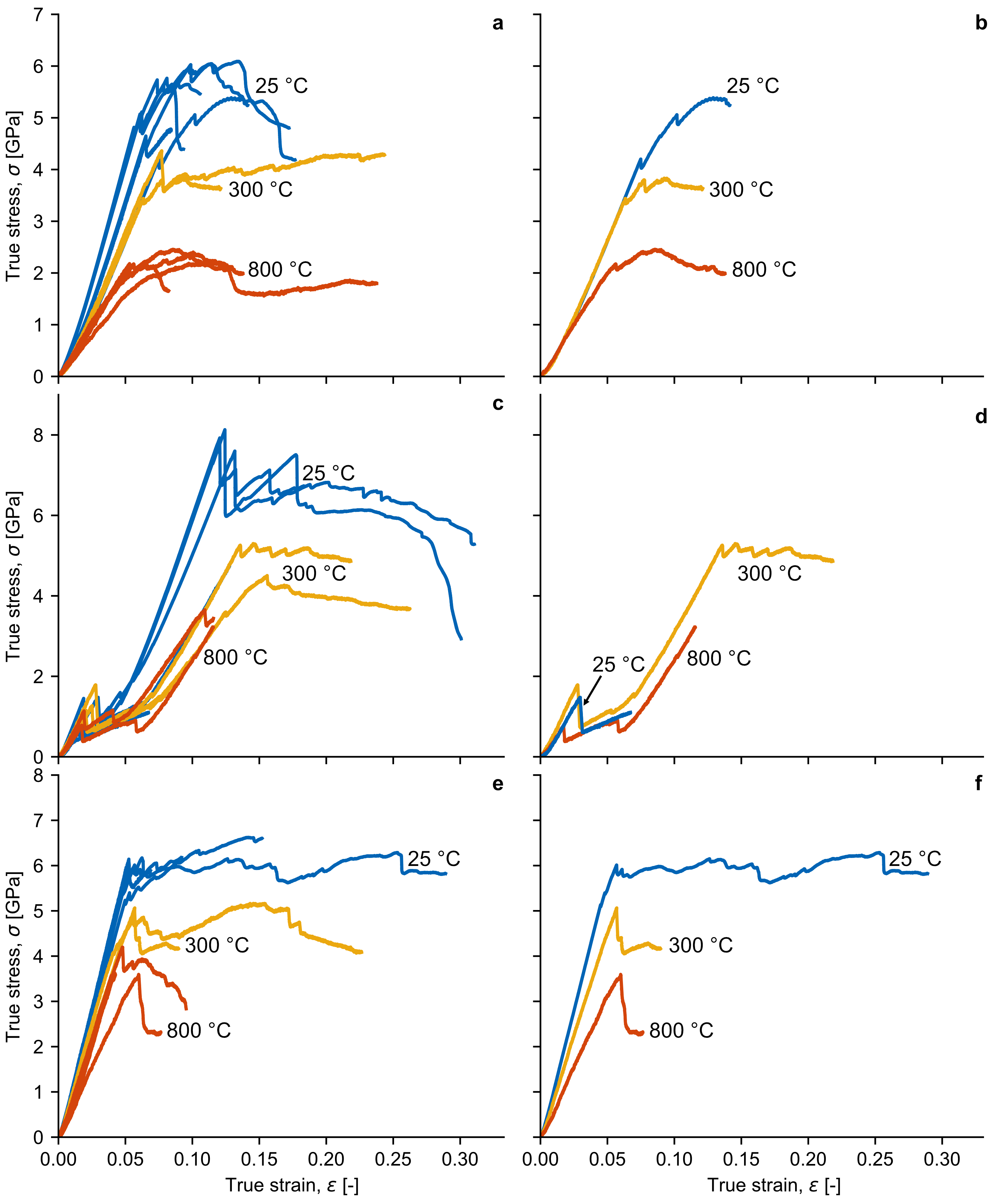} 
   \caption{True strain vs. true stress plots for orientation 1 pillars (a, b), orientation 2 pillars (c, d), and for orientation 3 pillars (e, f). Mechanical data that belongs to pillars, selected for nanostructural analysis, are exhibited in b, d, and f. Tests conducted at 25, 300, and 800 $^{\circ}$C are highlighted in blue, yellow, and red, respectively. a, b) Micropillar compression tests for orientation 1. }
   \label{StressStrain}
\end{figure}

Overall the reproducibility of the mechanical data is high (Fig. \ref{StressStrain}; Table \ref{ExpCond}). For deformation of pillars of orientations 1, we observe a scatter in yield stress that decreases with temperature from 7 to around 4\%. Scatter of measured yield stresses for pillars of orientation 2 also exhibits a decrease from 24 to 19\% but remains overall much higher than the yield-stress scatter of pillars of orientation 1. The scatter of the mechanical data for pillars of orientation 3 ranges between 8 and 11\% and shows no observable trend with changes in temperature (Table \ref{ExpCond}). 

\begin{figure}[h] 
   \centering
   \includegraphics[width=8cm]{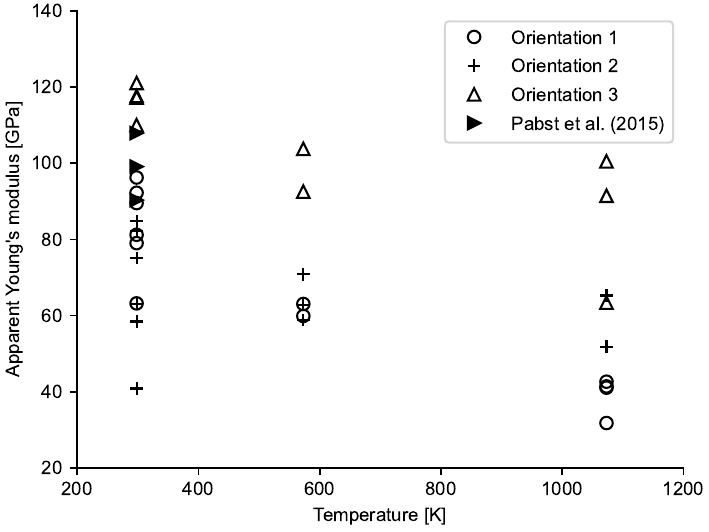} 
   \caption{Temperature vs. apparent Young's modulus for the three orientations together with published data on anorthite elasticity \cite{Pabst2015}. The three different values for anorthite }
   \label{E}
\end{figure}

Care must be taken when deducing Young's modulus \emph{E} from micropillar compression tests, because compliance of the testing device, surface flatness, alignment and any impurities on the diamond flat punch will all influence the mechanical data. We calculated the apparent Young's moduli for each test and observe a large scatter within each set of crystal orientation that appears to decreases with increasing temperature (Fig. \ref{E}). Despite the scatter, the calculated elastic moduli for pillars of orientations 1 and 3 deformed at 25 $^{\circ}$C are in good agreement with literature data on anorthite elasticity \cite{Pabst2015, Brown2016}.  Pillars of orientation 2 exhibit the widest spread in data at 25 $^{\circ}$C and are consistently more compliant than the literature data. The two outliers that exhibit an apparent Young's modulus of around 40 or 58 GPa, respectively, reveal an overall different mechanical behaviour, which probably results from off-centre pillar-punch loading geometries. Within the temperature range selected, i.e., between 25 and 800 $^{\circ}$C, we observe an overall softening of the elastic modulus with increasing temperature of around 25 to 50\% (Fig. \ref{E}). 
Please note, we neither corrected for the pillars indenting into the substrate nor for pillar taper. Since we are mainly interested in the post-yield behaviour of the pillars, and the calculated apparent Young's moduli reasonably agree with the literature data \cite{Pabst2015, Brown2016}, we refrained from correcting the elastic part of the deformation.

\begin{figure}[h] 
   \centering
   \includegraphics[width=8cm]{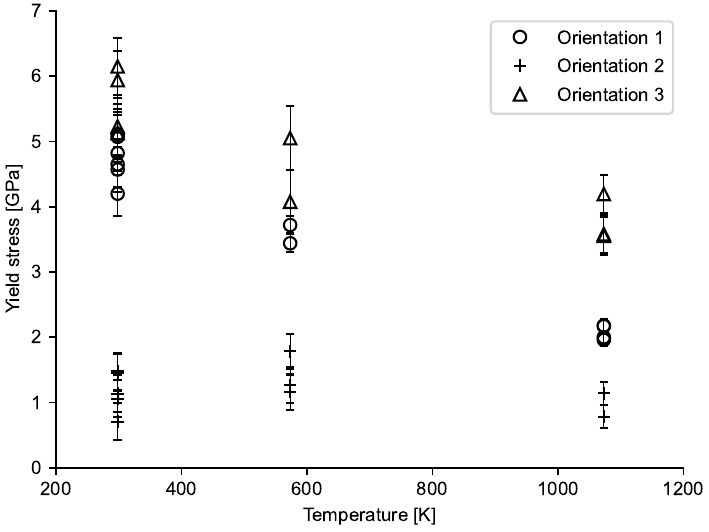} 
   \caption{Temperature vs. yield stress for the three different orientations.}
   \label{yield}
\end{figure}

Except for pillars of orientation 2, which start to yield at strains of around 1 to 2\% (Fig. \ref{StressStrain}c, d), all other pillars begin to plastically yield at higher strains of around 5 to 6\% (Fig \ref{StressStrain}a, b, e, f). Yield stresses for pillars of orientation 2 do not change with temperature whereas pillars of orientation 1 reveal a linear decreases of yield stress with increasing temperature (Fig. \ref{yield}). Yield stresses of pillars of orientation 3 appear to be less sensitive to changes in temperature than pillars of orientation 1. Regardless of temperature, pillars of orientation 3 are consistently stronger than pillars of the two other orientations, and pillars of orientation 2 are always weakest (Fig. \ref{yield}). 

\begin{figure}[htbp] 
   \centering
   \includegraphics[width=8cm]{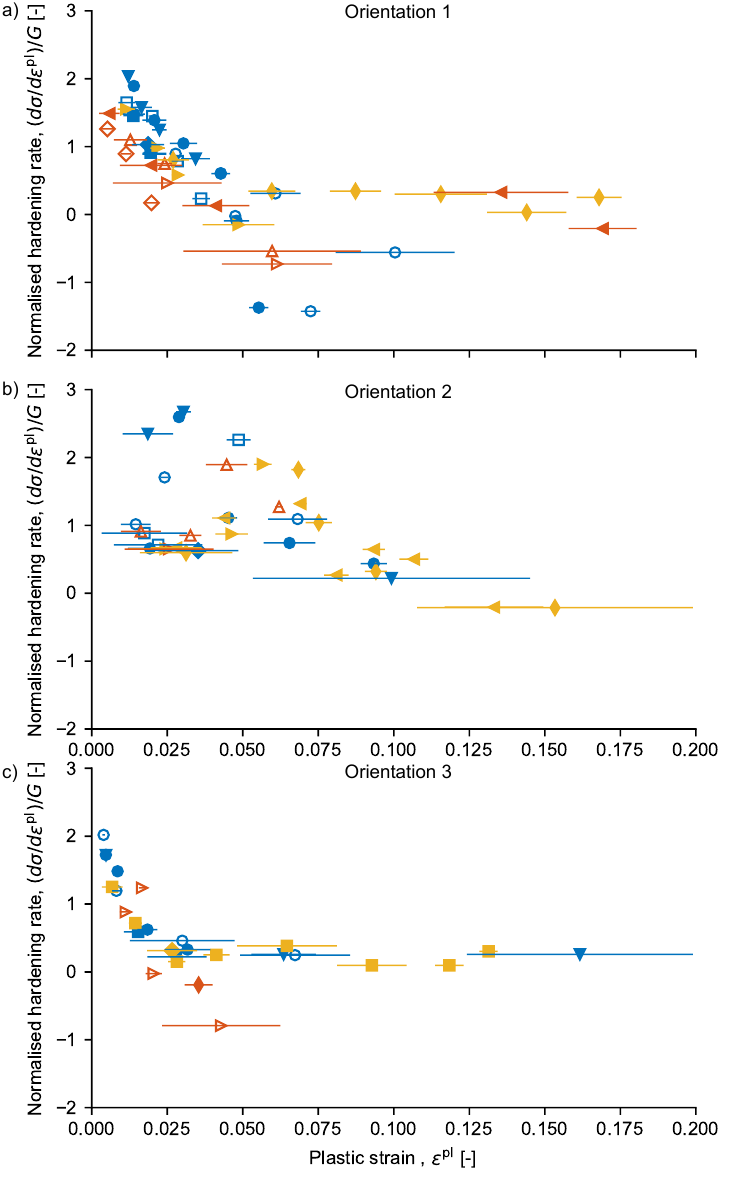} 
   \caption{Plastic strain vs. normalised hardening rate. a) Normalised hardening rate for orientation 1 pillars (a), orientation 2 pillars (b) and orientation 3 pillars (c). Blue, yellow, and red symbols refer to the temperature during testing, i.e., either 25, 300, or 800 $^{\circ}$C, respectively. Different symbols correspond to different tests, and the horizontal lines highlight the range of plastic strain over which the hardening rate was calculated.}
   \label{hardening}
\end{figure}

The overall post yield-point behaviour is considerably different between the three sets of orientations and changes with temperature within each set. We calculated the shear modulus \emph{G} for each pillar by using the relation between elastic and shear modulus \begin{math} G = E/(1 + \nu)\end{math}, and a constant poisson's ration of $\nu$= 0.295 (Pabst et al. 2015), to then calculate a normalised hardening rate for each deformation test (Fig. \ref{hardening}). The calculated shear moduli range from and are thus similar to the isotropic shear modulus for anorthite with An$_{96}$ \cite{Brown2016}\\
We observe that for pillars of orientation 1, the hardening rate (between 0 and 5\% plastic strain) appears to depend on temperature, with a decrease in overall hardening rate with increasing temperature (Fig. \ref{hardening}a). Pillars deformed to larger strains exhibit strain hardening behaviour up until around 12\% strain beyond which we observe the onset of strain weakening -- the continuous decrease of stress with strain -- which is as well reflected by the normalised hardening rate (Fig. \ref{hardening}a). During deformation at 300 $^{\circ}$C, the pillar deformed to a strain of around 12\% (O1P8) reveals post-yield hardening followed by strain weakening whereas the pillar, which was deformed to a higher strain of around 24\% (O1P7) exhibits overall strain hardening until we stopped the test. The four tests on pillars deformed at 800 $^{\circ}$C exhibit post-yield hardening and subsequent weakening that takes place at different strains (Figs. \ref{StressStrain}a, b; \ref{hardening}a). The pillar deformed to the highest strains, i.e., around 23\%, exhibits re-strengthening until the end of the test after initial softening at around 12\% strain (Fig. \ref{StressStrain}a). \\
At the lowest temperature of 25 $^{\circ}$C, pillars of orientation 1 exhibit stress drops of similar magnitudes of a few hundred MPa (Fig. \ref{StressStrain}a, b). Besides one larger stress drop that took place at the onset of yield at 300 $^{\circ}$C in pillar O1P7, stress drops are only a few tens of MPa and thus smaller in magnitude relative to those recorded in pillars deformed at 25 $^{\circ}$C (Fig. \ref{StressStrain}c, d). Overall, the mechanical curves for tests on pillars deformed at 800 $^{\circ}$C appear relatively smooth as we only detected two clear stress drops in the four runs, which were each a few tens of MPa large (Fig. \ref{StressStrain}e, f). 

The deformation behaviour of pillars of orientation 2 is significantly different to the evolution of stress with strain for pillars of the two other orientations. The first stage of deformation, up to strains of around 6\%, is very similar for all pillars of orientation 2 regardless of temperature (Fig. \ref{StressStrain}c, d). Yield of all pillars is accompanied by a stress drop of several hundreds of MPa. Subsequently, pillars deformed at 25 and 300 $^{\circ}$C exhibit an increase in stress up until a strain of around 4 to 6\%. Regardless of temperature, hardening rate appears to be overall constant at around 0.7 (Fig. \ref{hardening}b). This hardening stage at low strains is often accompanied by small stress drops that are generally only a few tens of MPa large (Fig. \ref{StressStrain}; Fig. S1). At low strains, the only difference is that pillars deformed at 25 and 300 $^{\circ}$C exhibit only one large stress drop at around 1 to 2\% strain while both pillars deformed at 800 $^{\circ}$C reveal two stress drops, the first at around 2\% and a second between 4 and 6\% strain (Fig. \ref{StressStrain}c, d). In contrast to the stress drops of pillars deformed at 25 and 300 $^{\circ}$C, the two stress drops in the test runs at 800 $^{\circ}$C are slightly smaller in size but both of equal magnitude. \\
Beyond around 6\% strain, stress rapidly increases in every test to either around 8 GPa (25 $^{\circ}$C), around 4.2 GPa (300 $^{\circ}$C) or to around 3.7 GPa (800 $^{\circ}$C) resulting in concave shaped stress-strain curves (Fig. \ref{StressStrain}c, d). Pillars deformed at 25 $^{\circ}$C then exhibit a sequence of large stress drops (up to 1 or 2 GPa) until stress fluctuates around a constant value of around 6.5 GPa before it eventually starts to either decreases slowly (O2P3) or rapidly (O2P2) beyond strains larger than around 22\%. Both pillars deformed at 300 $^{\circ}$C either reveal a sequence of equally large stress drops at a stress of around 5 GPa (O2P9) or a few stress drops followed by continuously decrease in stress until the end of the test (O2P7; Table \ref{ExpCond}). Tests conducted at 800 $^{\circ}$C were stopped at low strains of around 10\%, at which one pillar reveals a small stress drop of a few hundred MPa. 

Pillars of orientation 3 deformed at 25 and 300 $^{\circ}$C exhibit a sequences of strain hardening and weakening (Fig. \ref{StressStrain}e, f). When deformed to larger strains of around 23 or 28\%, pillars reveal that, in general, stresses fluctuate around either 5.9 GPa (25 $^{\circ}$C) or 4.8 GPa (300 $^{\circ}$C; Fig. \ref{StressStrain}e), and that this fluctuation is qualitatively stronger in the pillar deformed at 300 $^{\circ}$C (O3P3) than at 25 $^{\circ}$C (O3P5; Fig. \ref{StressStrain}e). At 800 $^{\circ}$C, we observe that pillars seem to immediately weaken after yield (Fig. \ref{hardening}c). In contrast to pillars of orientation 1, revealing a transition from overall hardening to weakening with increasing temperature to larger strains, the opposite is true for pillars of orientation 3 (Fig. \ref{hardening}); Pillars deformed at 800 $^{\circ}$C exhibit immediate weakening after yield, whereas pillars deformed to large strains at lower temperatures of either 25 or 300 $^{\circ}$C never exhibit overall weakening. This observations may however be biased as we only conducted two tests at 800 $^{\circ}$C and stopped those runs at low strains ($<$10\%). Furthermore, in contrast to the data on deformation of pillars of orientation 1, we observe that the stress-drop size appears to increase with temperature and that the stress drops, occurring during deformation at either 300 or 800 $^{\circ}$C, appear to not consist of a single event but are rather the sum of several events (Fig. \ref{StressStrain}f).

\begin{figure}[htbp] 
   \centering
   \includegraphics[width=8cm]{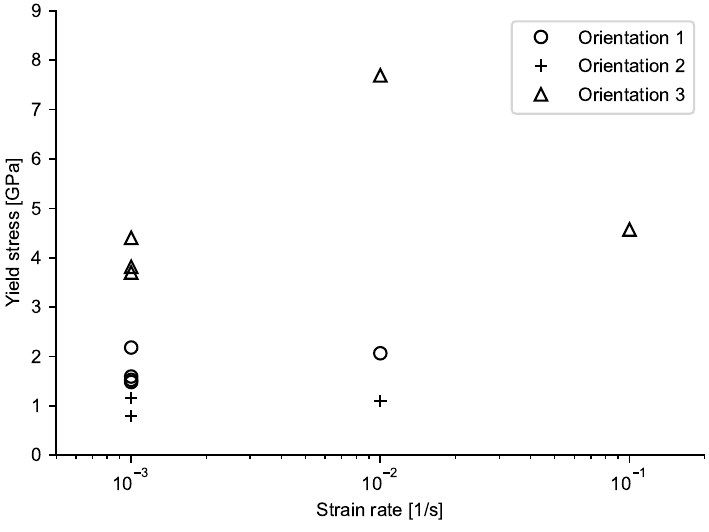} 
   \caption{Strain rate vs. yield stress plot revealing that there is no detectable influence of strain rate on yield strength for pillars of any orientation. The only test that shows higher stresses belongs to a pillar also revealing a significantly higher stiffness (Fig. \ref{StrainRate2}). All tests were conducted at 800 $^{\circ}$C. See text for further information.}
   \label{StrainRate}
\end{figure}

\begin{figure}[htbp]
	\centering
	\includegraphics[width=8cm]{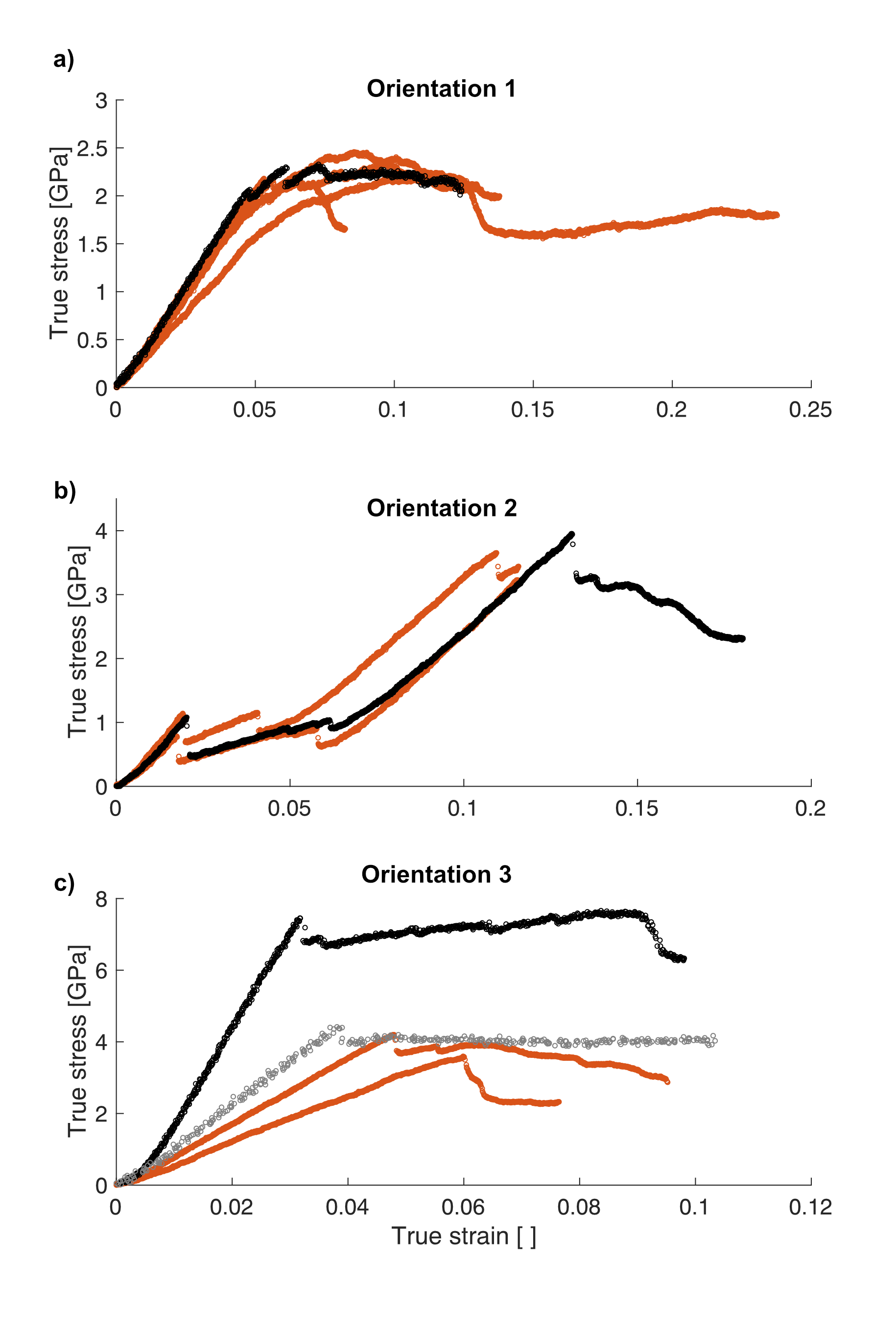}
	\caption{True strain vs. true stress curves for tests conducted under 800 $^{\circ}$C on pillars of orientation 1 (a), orientation 2 (b), and orientation 3 (c). Red curves= strain rate of 10$^{-3}$ 1/s; black curves=  strain rate of 10$^{-2}$ 1/s; grey curve= strain rate of 10$^{-1}$ 1/s.}
	\label{StrainRate2}
\end{figure}

In addition to tests performed with a strain rate of 10$^{-3}$ s$^{-1}$, we deformed pillars with either 10$^{-2}$ or 10$^{-1}$ s$^{-1}$ to test the strain rate sensitivity of the deformation mechanism controlling yield stress or strength for each pillar orientation. In this set of experiments, we kept the temperature constant at 800 $^{\circ}$C. Besides one test, conducted on a pillar of orientation 3 with a strain rate of 10$^{-2}$ s$^{-1}$ that is roughly twice as strong as pillars deformed with a strain rate of 10$^{-3}$ s$^{-1}$, we do not observe any noticeable variation in strength over the tested range in strain rate from 10$^{-3}$ to 10$^{-1}$ s$^{-1}$ (Fig. \ref{StrainRate}). The higher strength of pillar O3P10 can, however, be entirely explained by its higher stiffness (Table \ref{ExpCond}; Fig. \ref{StrainRate2}), which is potentially due to a different loading geometry.

\subsection{Nanostructural investigation}
\begin{figure}[htbp] 
   \centering
   \includegraphics[width=8cm]{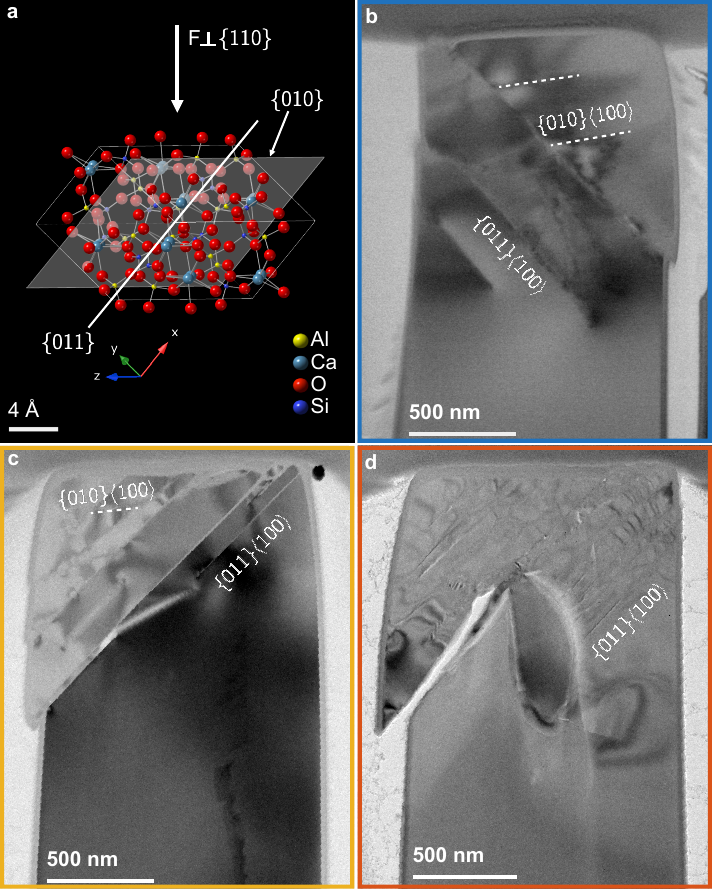} 
   \caption{Nanostructural analyses of pillars of orientation 2. a) Direction of force (F) with respect to the crystal orientation. The force is perpendicular to \{110\}. \{011\} is highlighted as solid white line, and \{010\} as transparent plane. b) Pillar deformed at 25 $^{\circ}$C revealing two different sets of slip bands, i.e., \{011\}$\langle$100$\rangle$ and \{010\}$\langle$100$\rangle$ (O1P1; Table \ref{ExpCond}). c) \{011\}$\langle$100$\rangle$ and \{010\}$\langle$100$\rangle$ slip bands in a pillar deformed at 300 $^{\circ}$C (O1P8; Table \ref{ExpCond}). d) Only \{011\}$\langle$100$\rangle$ slip bands are visible in the nanostructure of a pillar deformed at 800 $^{\circ}$C (O1P10; Table \ref{ExpCond}).}
   \label{TEM1}
\end{figure}

\begin{figure}[htbp] 
   \centering
   \includegraphics[width=8cm]{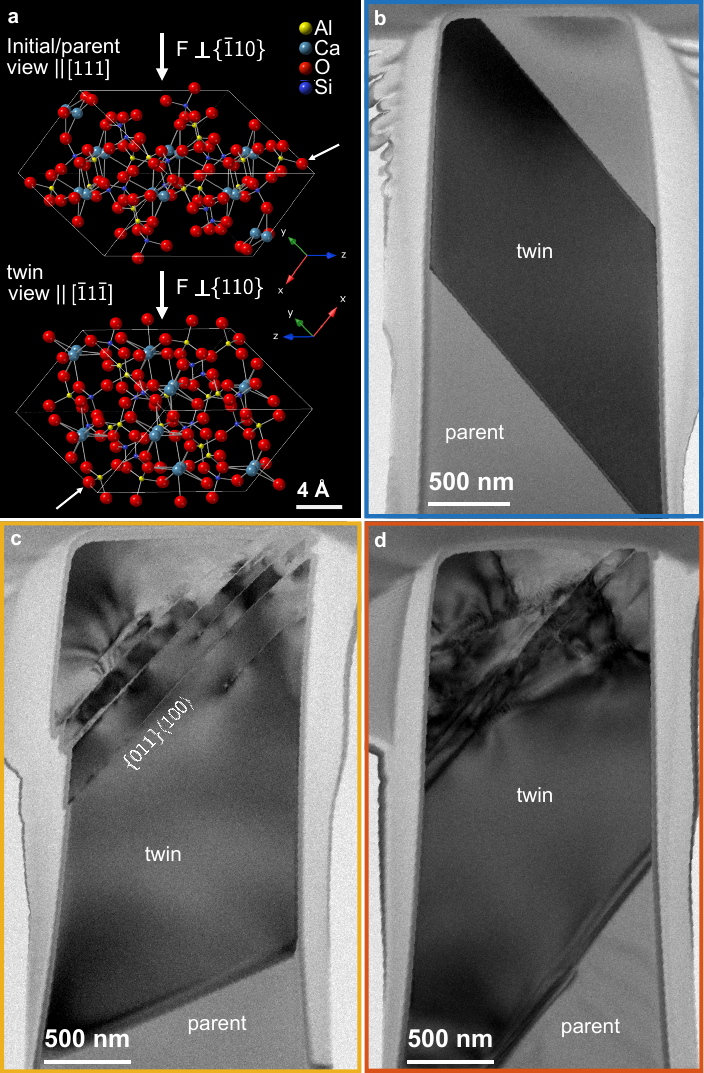} 
   \caption{Nanostructural analyses of pillars of orientation 2. a) Direction of force (F) with respect to the crystal orientation. The force is perpendicular to $\{\bar{1}10\}$ for the initial crystal orientation (parent orientation) or perpendicular to \{110\} for the twin. b) Deformation twin and adjacent parent in a pillar deformed under 25 $^{\circ}$C (O2P6; Table \ref{ExpCond}). c) Twin and parent in a pillar deformed under 300 $^{\circ}$C (O2P9; Table \ref{ExpCond}). Slip bands formed in twinned volume. d) Twin and parent in a pillars deformed under 800 $^{\circ}$C (O2P11; Table \ref{ExpCond}). }
   \label{TEM2}
\end{figure}

\begin{figure}[htbp] 
   \centering
   \includegraphics[width=8cm]{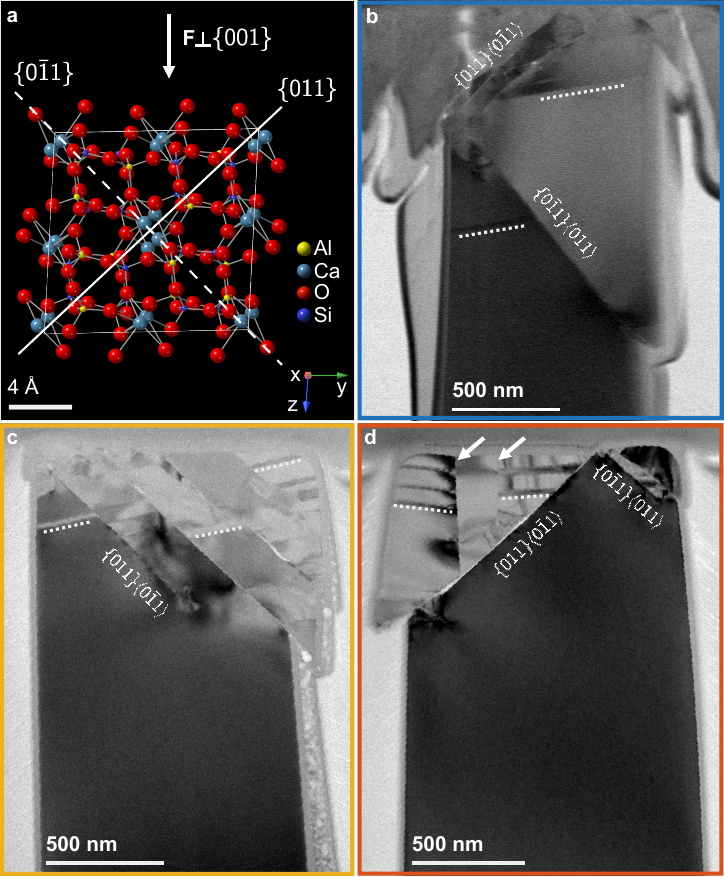} 
   \caption{Nanostructural analyses of pillars of orientation 3. a) Direction of force (F) with respect to the crystal orientation. The force is perpendicular to \{001\}. Active slip planes are indicated by either solid or dashed white lines. b) Cross-sectional view of a pillar deformed at 25 $^{\circ}$C (O3P3; Table \ref{ExpCond}) exhibiting slip bands and pericline twins (white dashed lines). c) Nanostructure of a pillar deformed at 300 $^{\circ}$C showing several slip bands (O3P6; Table \ref{ExpCond}). The white dashed light highlight some pericline twins. d) Pillar deformed at 800 $^{\circ}$C revealing slip bands and mainly twins (pericline= dashed white lines; albite= white arrows; O3P7; Table \ref{ExpCond}). All images were taken at the TEM and all pillars were deformed with a strain rate of 10$^{-3}$ s$^{-1}$}
   \label{TEM3}
\end{figure}

\begin{figure}[htbp]
	\centering
	\includegraphics[width=13cm]{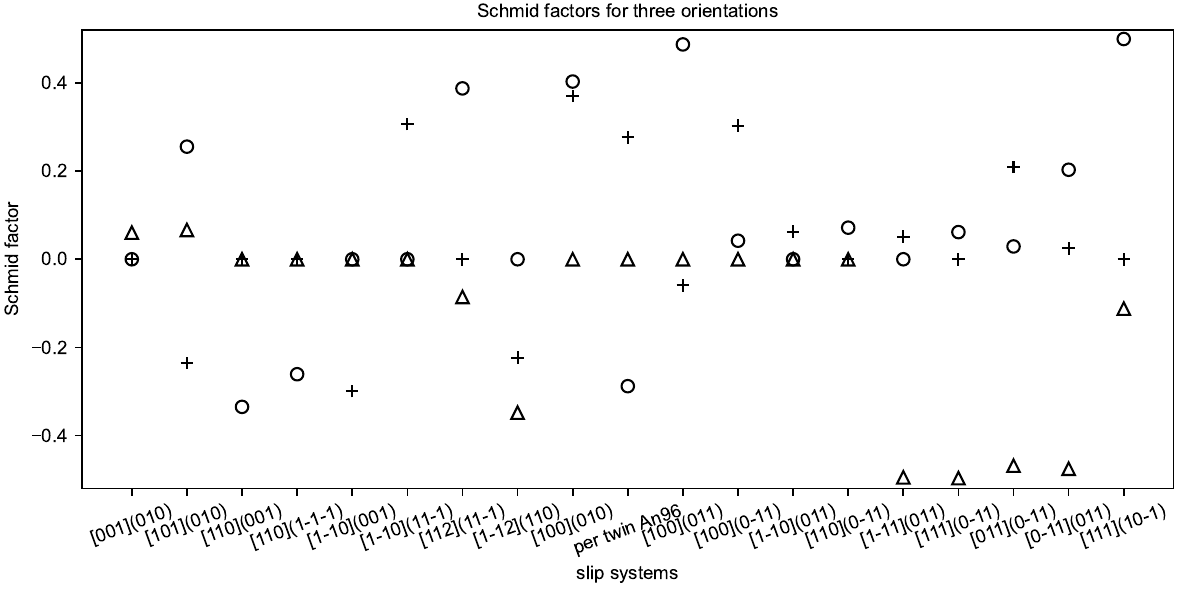}
	\caption{Calculated Schmid factors. See text for further explanations.}
	\label{Schmid}
\end{figure}

Selected area diffraction of the undeformed pillars reveals that the applied force \emph{F} was perpendicular to (i) \{110\} for orientation 1 pillars, (ii) \{$\bar{1}$10\} for orientation 2 pillars, and (iii) \{001\} for orientation 3 pillars (Figs. \ref{TEM1}a; \ref{TEM2}a; \ref{TEM3}a). To identify the active slip system(s), we first calculated the resulting Schmid factor \emph{m} (\begin{math} m = cos(\lambda) \cdot cos(\phi)\end{math}, with $\lambda$ and $\phi$ being the angles between the force and the slip plane or direction, respectively) for plagioclase slip systems found in natural samples (Fig. \ref{Schmid}; \citet{Olsen1984a, Olsen1985, Ji1990, Mehl2008, Miranda2016}. We then checked which slip systems with high Schmid factors match the nanostructural observations (Figs. \ref{TEM1}; \ref{TEM2}; \ref{TEM3}). Burgers vectors indicated are therefore based on calculated Schmid factors as well as on energy considerations \citep{Weertman1964}.

All investigated nanostructures that correspond to orientation 1, belong to pillars, which were strained to a similar total strain of around 14\% (Figs. \ref{TEM1}b-d; \ref{StressStrain}b). In those nanostructures, we found that \{011\}$\langle$100$\rangle$ with an \emph{m} of around 0.49 was mainly active, followed by the activation of \{010\}$\langle$100$\rangle$ with an \emph{m} of around 0.40 (Fig. \ref{Schmid}). We infer this sequence of slip-system activation as \{010\}$\langle$100$\rangle$ shows no cross-cutting relationship with \{011\}$\langle$100$\rangle$ but solely forms slip bands bound by \{011\}$\langle$100$\rangle$ slip bands (Fig. \ref{TEM1}). In images taken in TEM bright field (BF) mode, we sometimes observe strong and complex contrast variations where both slip systems interact (Fig. \ref{TEM1}b). It appears that the number of \{011\}$\langle$100$\rangle$ slip bands increases with increasing temperature from only three slip bands visible after deformation at 25 $^{\circ}$C, to around seven slip bands after deformation at 300 $^{\circ}$C, to more than 10 slip bands after deformation at 800 $^{\circ}$C (Fig. \ref{TEM1}b-d). At the same time, the number of slip bands that belong to \{010\}$\langle$100$\rangle$ seems to exhibit a negative correlation with temperature. In the pillar deformed at 25 $^{\circ}$C we can identify several \{010\}$\langle$100$\rangle$ slip bands, whereas we observe only a few in the pillar deformed at 300 \{010\}$\langle$100$\rangle$, and none in the pillar deformed at 800 $^{\circ}$C, although all three pillars were deformed to a similar total strain. In the central part of the pillar deformed at 800 $^{\circ}$C, we observe a crack that is around 500 nm long and exhibits an inclination of around 60$^{\circ}$ towards the direction of compression (Fig. \ref{TEM1}d). It is unclear whether this crack was present initially or produced during the test. 

The investigated pillars of orientation 2 were all deformed to different strains of around 7, 12, or 23\%, respectively (Figs. \ref{StressStrain}d). Regardless of total strain and temperature, all pillars exhibit a band, appearing dark in BF mode, indicating a difference in orientation between the brighter and the darker parts of the pillars (Fig. \ref{TEM2}). We compared the SAED of the initial pillar orientation with those taken from the dark-appearing bands and found that their corresponding zone axis (i.e, view or beam direction) can be transformed into one another by a rotation of 180$^{\circ}$ around [010] (Fig. \ref{TEM2}a) -- the transformation corresponding to the pericline twin law \cite{Nespolo2025}. Hence, the dark-appearing bands are pericline twins and the remaining brighter parts are the un-twinned volume or parent. A rotation of 180$^{\circ}$ about [010] results in the transformation of the pillars' initial orientation from \emph{F} perpendicular to $\{\bar{1}10\}$ to \emph{F} perpendicular to \{110\}, with the latter being the initial orientation of pillars of orientation 1 (Figs. \ref{TEM2}; \ref{TEM1}). Consequently, the twinned volumes ended up with the same orientation as pillars of orientation 1, and thus the slip bands found in the twinned volume were identified to be the result of slip along \{011\}$\langle$100$\rangle$ (Fig. \ref{TEM2}c).  

Nanostructural analyses of orientation 3 pillars reveal that deformation was accommodated by several mechanisms (Fig. \ref{TEM3}). After deformation at 25 and 300 $^{\circ}$C, respectively, we observe conjugated slip along two planes that exhibit intersection angles of either around 89 or 91$^{\circ}$. Together with the information on the initial orientation of the pillars, we identified these planes as \{011\} and $\{0\bar{1}1\}$, respectively. Since slip must occur within each plane and generally occurs along short Burger's vectors due to energy considerations \citep[e.g.,][]{Weertman1964}, we suggest $\langle0\bar{1}1\rangle$ or $\langle$11$\rangle$ as slip directions. Furthermore, these slip systems are with \emph{m}= 0.48 favourably oriented for slip and reveal the expected intersection angle of around 90$^{\circ}$ (Fig. \ref{TEM3}b, d; Fig. \ref{Schmid}). In addition to deformation being accommodated by slip, we observe pericline twinning in all pillars regardless of temperature (dashed lines in Fig. \ref{TEM3}b, c). The extent of pericline twinning appears to increase with increasing temperature (Fig. \ref{TEM3}b-d). In addition to pericline twins, we find an albite twin in the nanostructure of the pillar deformed at 800 $^{\circ}$C (white arrows in Fig. \ref{TEM3}d). 

\section{Discussion}

Before discussing the experimental and nanostructural results in more detail, we find it important to stress the fact that deformation of all pillars was predominantly accompanied by plastic deformation -- even when deformed under ambient temperature with strain rates of 10$^{-3}$ s$^{-1}$. Only a few pillars, deformed at 25 $^{\circ}$C and 10$^{-3}$ s$^{-1}$ failed in a brittle manner at the highest strains, i.e., $>$15\% strain for pillars of orientation 1 and $>$30\% strain for pillars of orientation 2. The overall dominance of plastic deformation is somewhat surprising for plagioclase is known to easily deform by cleavage cracking \citep[e.g.,][]{Tullis1992, Stunitz2003, Incel2024} or has been found to be prone to amorphisation at elevated stresses \citep[e.g.,][]{Yund1990, Pec2012a, Pec2021, Incel2025}. Especially amorphisation of anorthite was an initial concern as recent micropillar compression tests on olivine revealed partial amorphisation along specific planes \cite{Kranjc2020}. We can therefore state that micropillar compression is suited to study and quantify plastic deformation in plagioclase.

\subsection{Strength of anorthite micropillars}
Since anorthite is a mechanically anisotropic material \citep{Brown2016}, it is not surprising that our pillars reveal differences in strength depending on orientation (Fig. \ref{yield}). Pillars of orientation 2, which were intentionally oriented to activate twinning, indeed exhibit the onset of plastic deformation by twin nucleation and growth, confirmed by the nanostructural analyses (Fig. \ref{TEM2}). The twins that formed obey the pericline twin law, i.e., a rotation of 180$^{\circ}$ around [010] \citep[e.g.,][]{Nespolo2025}. Hence, the strength of pillars of orientation 2 is controlled by the nucleation of pericline twins. The latter appears to be insensitive to changes in either temperature or strain rate (Figs. \ref{yield}; \ref{StrainRate}), which is expected for twinning \cite{Christian1995}. Twin nucleation is associated with the collective motion of atoms along specific crystallographic planes that takes place at speeds high enough to produce acoustic emissions. \citet{Laughner1979} have recorded acoustic emissions due to Dauphiné twinning in quartz, and tin can produce audible sounds during twinning, called tin cry  \citep[e.g.,][]{Chalmers1932}. Because of this inherently rapid nature of the twinning process, the range in commonly applied strain rates is generally too low to exert a noticeable influence of strain rate on twin nucleation. Due to the underlying collective character of atoms moving to produce a twin, temperature has no strong influence on twin nucleation either, in contrast to the glide of dislocations that must overcome energy barriers, e.g., lattice resistance, precipitates, other dislocations. 

It is less obvious to determine the strength-controlling mechanism for pillars of orientation 1 and 3. Small volumes may behave quite differently to their larger counterparts as the former either contain fewer initial defects or no initial defects at all, and dislocations could rapidly leave the small pillars at the free surfaces \citep[e.g.,][]{Uchic2004, Greer2005}. Almost all micropillar studies have been conducted on high symmetry materials, e.g., fcc or bcc metals or ceramics. Since there are already reported differences in micromechanical behaviour between fcc and bcc metals \citep[e.g.,][]{Christian1983, Rizzardi2022} as well as between metals and ceramics \citep[e.g.,][]{Korte2011}, a comparison of our results with the existing data, especially with data on ceramics that share several properties, e.g., high lattice resistance, large unit cells, is clearly insightful, but requires caution due to the fundamental structural difference between these materials. 

The clear temperature sensitivity of strength observed in pillars of orientation 1 and 3, which is stronger for pillars of orientation 1 (Fig. \ref{yield}), demonstrate that the underlying mechanism controlling yield strength is limited by a thermally activated barrier (Fig. \ref{yield}). Whether the strength-controlling process is dislocation nucleation/multiplication or the resistance of the lattice to dislocation glide, is not conclusive at this point, because both are sensitive to changes in temperature and we find good arguments for either process in the collected data. Since most slip bands are connected with the pillar surfaces (Fig. \ref{TEM1}; \ref{TEM3}), it is possible that strength is controlled by a coupled process of dislocation nucleation at the surface and subsequent glide, which could be summarised as dislocation-source activity, i.e., the ability to nucleate dislocations and their glide away from the source. Changing the strain rate by only one to two decades may have been too small to highlight any significant strain rate sensitivity; Especially if the strength-controlling mechanism is associated with a large activation volume, e.g., collective motion of dislocations (see Fig. 4b in \citet{Jennings2011}. We can, however, exclude a stochastic activation of pre-existing defects in the pillars, which should change in number, length and orientation from pillar to pillar. If strength was controlled by the internal defect geometry, we would expect a high scatter in recorded yield stresses, which does not fit the observed high reproducibility of the data (Table \ref{ExpCond}; Fig. \ref{StressStrain}). 

\subsection{Post-yield deformation}

We find further evidence for the operation of thermally activated deformation mechanisms in the post-yield deformation behaviour of pillars of orientations 1 and 3. With increasing temperature, we observe an overall decrease in magnitude of the distinct stress drops detected in the deformation curves of pillars of orientation 1, which could either mirror the expected increase in source activity or facilitated glide as lattice resistance decreases with increasing temperature. Both processes would lead to an increased number of slip bands, here \{011\}$\langle100\rangle$, with increasing temperature, which we can confirm through the nanostructural analyses (Fig. \ref{TEM1}). In pillars of orientation 1 deformed under 25 and 300 $^{\circ}$C, we observe the activation of a secondary slip system, i.e., dislocation activity on \{010\}$\langle100\rangle$. The incremental increase in stress due to either source inefficiency or resistance to glide on a single slip system at 25 or 300 $^{\circ}$C resulted in the activation of this secondary slip system with a lower Schmid factor (Fig. \ref{Schmid}). When deformed at 800 $^{\circ}$C, dislocation multiplication/mobility on \{011\}$\langle100\rangle$ was sufficient to fully accommodate the imposed deformation, highlighted by the exclusive formation of numerous \{011\}$\langle100\rangle$ slip bands (Fig. \ref{TEM1}d). The high activity of a single source and/or the ease for dislocations to glide along \{011\}$\langle100\rangle$, produced a more homogeneous deformation of the pillars deformed under 800 $^{\circ}$C, which is also reflected by the overall smooth character of the deformation curves (Figs. \ref{TEM1}d; \ref{StressStrain}a, d). \\
It has been discussed in the literature that in low symmetry triclinic materials such as anorthite, or plagioclase in general, dislocations with a true edge character are not possible \cite{Stunitz2003, Olsen1984a, Montardi1987}. It therefore seems reasonable to assume that the observed increased mobility of mixed mode or pure screw dislocations may in fact be caused by the temperature sensitivity of kink-pair nucleation in anorthite. Furthermore, only screw dislocations are able to change slip planes thus to cross slip. It is therefore possible that we observe cross slip between \{011\}$\langle100\rangle$ and  \{010\}$\langle100\rangle$, since both slip systems share the same Burger's vector, i.e., $\langle100\rangle$. We speculate that cross slip could lead to the immobilisation of dislocations eventually resulting in the reduction of ductility and brittle failure of pillars deformed at 25 $^{\circ}$C to large strains (Figs. \ref{StressStrain}a, b; \ref{hardening}a). In these pillars cross slip should be most sufficient due to the highest detected number of \{010\}$\langle100\rangle$ (Fig. \ref{StressStrain}a). 

The post-yield deformation behaviours of pillars of orientation 3 are strikingly different to those observed for pillars of orientation 1. In contrast to pillars of orientation 1, we do not observe a clear increase in slip-band formation of the dominant slip systems with temperature, but instead an increase in the amount of twins present after deformation (Fig. \ref{TEM3}). The pillars deformed under either 300 or 800 $^{\circ}$C were strained by roughly the same amount, yet, the pillar deformed under 300 $^{\circ}$C exhibits a higher number of slip bands than the pillar deformed under 800 $^{\circ}$C, which in turn reveals a higher twin density (Fig. \ref{TEM3}c, d). The formation of twins in any of these pillars is particularly interesting, because no to little shear stress was resolved on them, indicating that the deformation of pillars of orientation 3 do not follow Schmid's law (Fig. \ref{Schmid}). In contrast to the stress drops recorded for pillars of orientation 1 and 2, stress drops recorded during deformation of pillars of orientation 3 appear to become larger with increasing temperature (Fig. \ref{StressStrain}). Furthermore, it seems that with increasing temperature each stress drop can no longer be attributed to one single event, but rather reflect a number of temporally closely spaced events (Fig. \ref{StressStrain}f), pointing towards one deformation mechanism triggering another. Combining all these observations, we speculate that there could be a link between dislocation mobility and twinning, e.g., an increased mobility of dislocations leading to stress concentrations that act as nucleation sites for twins.  

\begin{figure}[htbp] 
   \centering
   \includegraphics[width=8cm]{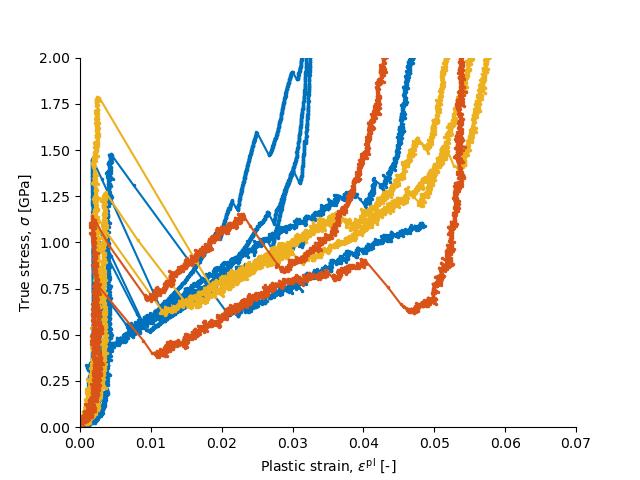} 
   \caption{Plastic strain vs. true stress for orientation 2. The jumps in plastic strain represent the nucleation of pericline twins.}
   \label{TwinPlastic}
\end{figure}

For we stopped deformation of one pillar of orientation 2 shortly after the initial stress drop but prior to the observed hardening at around 5 to 7\% strain (Fig. \ref{StressStrain}d), we know that the initial stress drop was caused by pericline twinning (Figs. \ref{StressStrain}d; \ref{TEM2}). To be precise, the stress drop is associated with twin nucleation whereas the subsequent hardening is linked with its growth. The latter is with a normalised hardening rate of 0.7 significantly higher than for twin growth in other materials (Fig. \ref{hardening}b; \citet[e.g.,][]{Prasad2014}. In order for the twins to grow, the stacking faults formed by twin disconnections much move through the lattice. Therefore, a possible reason for such elevated hardening rates could be related to difficulties in moving stacking faults through the anorthite lattice. As visible in Fig. \ref{TwinPlastic}, roughly 50 to 75\% of the total plastic twinning strain was accommodated by twin growth, whereas the remaining strain was produced by twin nucleation. We measured the twin thickness of pillar O2P5 to be around 1150 nm. Since this pillar experienced a plastic strain of around 4.8\%, a twinning shear $\gamma$ of around 0.11 would be necessary if plastic strain was completely accommodated by twinning, i.e., nucleation and growth together. Considering the structural differences between labradorite and anorthite, this calculated twinning shear matches well with the reported literature value of around 0.136 for pericline twinning in labradorite \cite{Xie2019}.

As discussed above, we neither observe an influence of temperature on twin nucleation nor on twin growth. However, temperature seems to affect the number of larger twins that form at low strains. During deformation under 25 and 300 $^{\circ}$C, we observe a first pronounced stress drop due to twin nucleation that are several hundreds of MPa large, sometimes followed by smaller stress drops of a few tens of MPa (Fig. \ref{StressStrain}c, d). When deformed under 800 $^{\circ}$C, however, we notice two stress drops that are both large, i.e., several hundreds of MPa. Since the magnitude of the stress drop correlates with twin thickness, the first twins that form under 25 and 300 $^{\circ}$C should be roughly 230 to 460 nm thick whereas the following twins would only be 20 to 50 nm wide. The two twins that formed during deformation under 800 $^{\circ}$C would be around 110 to 180 nm thick. The first twin that forms at around 1 to 2\% plastic strain is followed by a stage at which this newly formed twin grows. As strain due to twinning seems to saturate at around 4 to 5\% plastic strain or 5 to 6\% true strain, the subsequent twins will therefore remain considerably thinner (Figs. \ref{StressStrain}c, d; Fig. \ref{TwinPlastic}). The observed limited amount of strain being accommodated by twinning is in agreement with the assumption that twinning generally contributes only little to plastic deformation, e.g., \citet{Qu2025} report a limit for deformation being accommodated by twinning in calcite of around 2 to 4\% (see Fig. 9d in \citet{Qu2025}. \\
Because no further deformation can be accommodated by twinning at strains beyond around 5\% plastic strain, the twinned pillars will be elastically loaded until another deformation mechanism is activated (Fig. \ref{StressStrain}c, d). In our study, we observe the activation of dislocation slip within the twinned volume (Fig. \ref{TEM2}b-d). Since pericline twinning produced a rotation of the twin of 180$^{\circ}$ around [010], the twinned volume in pillars of orientation 2 reveals the same orientation as pillars of orientation 1. Consequently the slip bands we observe within the twinned volume were produced by slip along \{011\}$\langle100\rangle$. The latter starts at a significantly higher yield stress than expected from the mechanical data on orientation 1 pillars (Fig. \ref{StressStrain}). This higher yield stress can be explained by a rotation of the twin boundary (Fig. \ref{TEM2}b,c,d). Since the orientation of the pericline twin boundary did not change before slip was activated (Fig. \ref{TEM2}b, d), its rotation must occur prior to the onset of slip. For pillar O2P10, deformed at 300 $^{\circ}$C, a clockwise rotation of around 25$^{\circ}$, deduced from TEM analysis (Fig. \ref{TEM2}), would result in a reduction of the Schmid factor from 0.48 to around 0.28.

Overall, we observe strength and post-yield behaviour strongly depend on orientation. This observation rules out any significant contribution to the observed deformation behaviour due to sample preparation, i.e., FIB milling, as this process was kept identical for each pillar. Furthermore, the substantial decrease in strength with temperature in pillars of orientation 1 and 3 together with twinning being an athermal process, demonstrate that the determined strengths are not due to the temperature dependence of the shear modulus. It has been shown that the electron beam can induce plasticity and thus change the deformation behaviour of micropillars. We ran tests with the beam switched off and observe no difference to pillars deformed under otherwise identical experimental conditions (Table \ref{ExpCond}).  

\subsection{Interaction of deformation mechanisms}
It is striking that not only does the yield strength change with orientation, which is expected from a mechanically anisotropic material, but the way strain is accommodated and localises appears to change substantially with variations in orientation as well. In pillars of orientation 1, we observe that strain becomes more distributed on numerous slip bands with increasing temperature (Fig. \ref{TEM1}). The expected transition from macroscopically localised to homogeneous deformation with increasing temperatures observed in cm-sized samples and rocks \cite[e.g.,][]{Paterson1958}, is thus reflected at the µm-scale in a single grain. In contrast, we do not observe such strain delocalisation with increasing temperature in pillars of orientation 3 (Fig. \ref{TEM3}). As discussed above, the post-yield deformation behaviour of pillars of orientation 3 may be the result of an interaction between dislocations and twin formation. For twins can act as dislocation barriers and thus decrease the effective grain size (dynamic Hall-Petch effect), twinning has been found to increase strength and ductility in several materials, i.e., twin induced plasticity or short TWIP \cite{Grssel2000}. Yet, due to the limitation in strain being accommodated by twinning, ductility remains only high if dislocation motion between twin boundaries is still possible. Since the volume between twins is small and lack evidence for dislocation activity (Fig. \ref{TEM3}), we thus assume that in our case the interplay between dislocation slip and twinning may in fact cause a reduction in ductility, at least in micrometre-sized anorthite grains. The increased twin density in the pillar after deformation under 800 $^{\circ}$C could thus explain the counterintuitive decrease in ductility with temperature (Figs. \ref{StressStrain}e, f; \ref{hardening}c). Consequently, not just the strength, but the overall way a polycrystalline anorthite-rich rock deforms and the magnitude of plastic deformation it can accommodate may significantly change depending on the evolution of grain orientations.

\subsection{Critical resolved shear stresses}
\begin{figure}[htbp] 
   \centering
   \includegraphics[width=8cm]{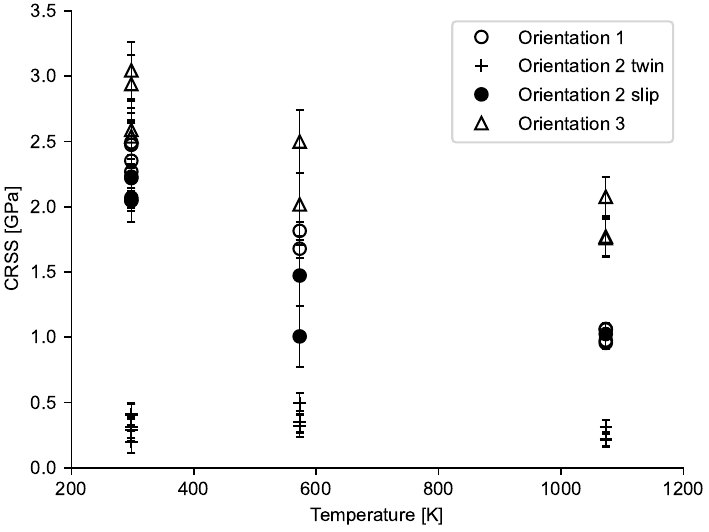} 
   \caption{Calculated CRSS for twinning (orientation 2; crosses), glide along \{011\}$\langle0\bar{1}1\rangle$ or $\{0\bar{1}1\}\langle011\rangle$ (orientation 3, triangle), or glide along \{011\}$\langle$100$\rangle$ (orientation 1, circles; orientation 2, filled circles). }
   \label{CRSS}
\end{figure}

Combining the mechanical data, i.e., yield stresses, with the nanostructural information, i.e., identification of the active slip system controlling deformation, we were able to calculate the critical resolved shear stresses (CRSS) necessary to activate dislocation motion and twinning. In our set of active/controlling slip or twin systems, we find that pericline twinning is with around 300 MPa the easiest deformation mechanism to activate, followed by the activation of the slip system \{011\}$\langle$100$\rangle$. The latter remains consistently softer than \{011\}$\langle0\bar{1}1\rangle$ or $\{0\bar{1}1\}\langle011\rangle$ within the range in temperature tested. Interestingly, this difference in strength becomes more pronounced at 800 $^{\circ}$C (Fig. \ref{CRSS}). As discussed above, pillars of orientation 2 reveal the onset of slip along \{011\}$\langle$100$\rangle$ at higher strains. Calculating the CRSS to activate \{011\}$\langle$100$\rangle$ in pillars of orientation 2 (Schmid factor \emph{m}= 0.28; filled circles in Fig. \ref{CRSS}), is in very good agreement with the CRSS for \{011\}$\langle$100$\rangle$ deduced from pillars of orientation 1 (\emph{m}= 0.49; open circles in Fig. \ref{CRSS}). 

Previous studies on plastic deformation in plagioclase discussed the importance of  [001](010) as slip system \citep[e.g.,][and references therein]{Olsen1984a, Olsen1985, Montardi1987, Ji1990, Stunitz2003, Mehl2008}. Pillars of the present study were, unfortunately, not favourably oriented to activate (010)[001] (Fig. \ref{Schmid}). Nevertheless, we believe that this slip system will play a less dominant role for crystal-plastic deformation in anorthite than it probably does in plagioclase. The latter generally reveals C$\bar{1}$ symmetry whereas our anorthite starting material shows a structure that results in P$\bar{1}$ symmetry. This reduction in symmetry involves the doubling of the c-axis from around 7 to 14 \AA. In anorthite, and in the absence of possible dislocation dissociation into partials along [001], slip along [100] with an associated Burger's vector of around 8.2 \AA would therefore be energetically much more favourable than along [001] with a Burger's vector of around 14.2 \AA, as the energy of a dislocation \emph{E} can be expressed as \emph{E}$\propto$\emph{b}$^2$ \cite{Weertman1964}, with \emph{b} being the Burger's vector. Please note that for pillars of orientation 1 and 3 these calculated CRSS are not necessarily linked to stresses required to overcome a Peierls barrier. Since we observe that dislocation nucleation may play a non-negligible role in these pillars, the stresses may rather reflect a critical stress to overcome dislocation nucleation and glide -- thus the birth of plasticity. 

The CRSS for twinning in calcite has been shown to depend on temperature as it decreases from around 10 MPa at ambient conditions down to just a few MPa under 800 $^{\circ}$C \cite{Turner1954, DeBresser1997}. Because twinning is insensitive to changes in strain rate, using twins as paleo piezometers is extremely helpful as one no longer needs to account for changes between laboratory and natural strain rate; The latter generally being several orders of magnitude slower than common strain rates used for laboratory experiments. In contrast to twinning in calcite, twinning in anorthite requires stresses that are around ten to even hundred times higher. Therefore, twinning in anorthite could already provide insights into the minimum stress levels necessary, i.e., a lower bound for external stresses, natural rocks experienced during deformation.

\subsection{Size effects \& extrapolation to nature}

\begin{figure}[htbp] 
   \centering
   \includegraphics[width=8cm]{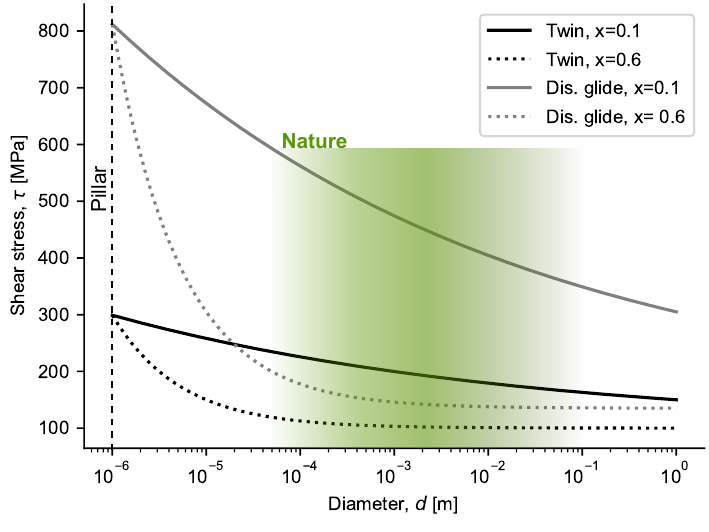} 
   \caption{Estimated evolution of strength with increase in size if size effect is strong, i.e., \emph{x}= 0.6, or weak, i.e., \emph{x}= 0.1 for twinning and dislocation glide, respectively. The green-shaded area covers the range in grain sizes of plagioclase in most lower crustal rocks. See text for further information.}
   \label{Size}
\end{figure}

Whilst the deformation of small volumes allows us to quantify dislocation-mediated deformation in hard and brittle materials, the obtained CRSS derived from micropillar compression cannot necessary be directly applied to natural rocks, which generally consist of grains with a few tens of micrometre to even a few centimetre in size. Previous studies reveal that the shear stress $\tau$, required to activate glide, decreases with increasing pillar size \emph{d} \citep[e.g.,][]{Uchic2004, Greer2005}, for the initial amount of dislocations that can be activated increases with pillar size. It has been shown that the relationship between shear stress and pillar size can be described by a simple power law of the form $\tau = \tau_{0}\emph{B}\emph{d}^{-x}$,  where $\tau_{0}$ represents the lattice resistance at 0 K with \emph{B} being a fitting parameter. Normalising $\tau$ by the shear modulus \emph{G} and the diameter by the Burger's vector \emph{b}, \citet{Dou2009} have shown that the power-law exponent \emph{x} is always around 0.6 regardless of the material used for the test. This universal relationship, however, only holds for fcc metals. \citet{Korte2011} demonstrated that micropillars of silicon, a metalloid with a high lattice resistance of around 1.4 GPa, exhibit a negligible size effect at low homologous temperatures, which the authors explain by the predominant contribution of lattice resistance to strength.
Extrapolating the determined shear stresses for dislocation motion on \{011\}$\langle100\rangle$ to 0 K, gives a lattice resistance $\tau_{0}$ for anorthite of around 2.7 GPa, similar to the reported lattice resistance for olivine of around 3 GPa \cite{Idrissi2016, Hansen2019, Wallis2026}. We therefore assume that at low homologous temperatures, the size of anorthite micropillars will have a similarly weak influence on strength. The estimated evolution of pillar strength with increasing pillar size for either weak or strong size effect, x= 0.1 or 0.6, respectively, is exhibited in Figure \ref{Size}. \citet{Borg1969} estimated a CRSS for twinning of around 100 MPa from deformation tests on millimeter-sized plagioclase single crystals under 0.8 to 1 GPa, 800 $^{\circ}$C, and strain rates of 10$^{-4}$ to 10$^{-5}$ s$^{-1}$. Please note that due to the activation of other deformation mechanisms besides twinning (see Table in \citet{Borg1969}, it is however unclear how the authors came to this conclusion. Stresses of at least 150 to 200 MPa would be required to activate twinning for millimetre to centimetre-sized grains when assuming a power-law exponent of \emph{x}= 0.1  (Fig. \ref{Size}). In contrast, choosing an exponent of \emph{x}= 0.6, a commonly determined power-law exponent for fcc metals \cite{Uchic2004, Greer2005, Dou2009}, the necessary stress for twinning would rapidly drop until around 10 µm beyond which it remained rather constant at around 100 MPa. By deforming stacked millimetre-sized single crystals of plagioclase under 900 $^{\circ}$C and a strain rate of around 8$\cdot$10$^{-7}$ s$^{-1}$, \citet{Stunitz2003} provide experimental data on the apparent strength of (010)[001]. Although we did not activate this particular slip system in our micropillar tests, and \citet{Stunitz2003} report a mixture of deformation mechanisms being active during their deformation tests, we use this data in the absence of better alternatives to get a preliminary idea on how strength would evolve with increasing sample diameter, if strength is controlled by dislocation glide. We used the same power-law exponents, \emph{x}= 0.1 and \emph{x}= 0.6, and the provided CRSS for dislocation glide on (010)[001] of around 135 MPa (Fig 2 in \citet{Stunitz2003}. Regardless of sample diameter, twinning remains weaker than dislocation glide in anorthite, which matches well with the observed abundance of deformation twins in natural plagioclase-rich rocks \citep[e.g.,][]{Mehl2008}. Despite these calculations being based on several more or less robust assumptions, we find this exercise quite insightful, as it gives us a first rough estimate on stress magnitudes necessary to activate the same process, e.g., twinning or dislocation nucleation and glide, in natural millimeter to centimeter-sized grains (Fig. \ref{Size}). Future tests on pillars with different diameters will not only reveal the size effect on strength in anorthite, but will help us decipher whether the mechanism controlling strength is the activity of dislocation sources (dislocation nucleation and glide) or lattice resistance (dislocation glide). A detailed investigation on how variations in size will affect the strength of anorthite micropillars will be the focus of a future study. 

\section{Conclusion}
Regardless of temperature, twinning is with a CRSS of around 300 MPa always the weakest deformation mechanism, followed by slip along \{011\}$\langle100\rangle$. Slip on \{110\}$\langle0\bar{1}1\rangle$ or $\{\bar{1}10\}\langle011\rangle$ is consistently the strongest mechanism in the temperature range tested. It is not conclusive yet, whether the calculated CRSS for slip reflect the stresses necessary for dislocation motion to overcome lattice resistance, i.e., Peierls stress, as dislocation nucleation at the pillar surfaces may play a non-negligible role. Overall pillar deformation and strain localisation appear to be strongly affected by changes in orientation, highlighting that the deformation behaviour of polycrystalline rocks may potentially differ depending on the evolution of grain orientations, i.e., texture. \\
Pericline twinning is an athermal process and insensitive to changes in strain rate, which could potentially make it a robust and helpful paleo piezometer; Especially because plagioclase is abundant in most crustal rocks. Since the CRSS for anorthite is at least one order of magnitude higher in anorthite than in calcite, twinning in plagioclase could already be useful as it helps differentiating between high-stress vs. low-stress regimes.\\
Micropillar compression is not only the method of choice when studying plastic deformation in hard and brittle ceramics, but has proven itself tremendously useful in quantifying plastic deformation of Earth materials as well. Since most of the rock-forming minerals are in fact silicates and thus generally hard and brittle, we believe that the use of this method now enables us to quantify processes and material properties that remained inaccessible using standard rock deformation techniques in the past. 

\pagebreak
 
\bibliographystyle{elsarticle-harv} 

\bibliography{pillar}

@article{Grssel2000,
   author = {O. Grässel and L. Krüger and G. Frommeyer and L. W. Meyer},
   journal = {International Journal of Plasticity},
   pages = {1391-1409},
   title = {High strength Fe±Mn±(Al, Si) TRIP/TWIP steels development - properties - application},
   volume = {16},
   url = {www.elsevier.com/locate/ijplas},
   year = {2000}
}

@article{Christian1983,
   author = {J.W. Christian},
   journal = {Metallurgica Transactions A},
   pages = {1237-1256},
   title = {Some surprising features of the plastic deformation of body-centred cubic metals and alloys},
   volume = {14A},
   year = {1983}
}

@article{Laughner1979,
   author = {J W Laughner and T W Cline and R E Newnham and L E Cross},
   journal = {Physics and Chemistry of Minerals},
   pages = {129-137},
   title = {Acoustic Emissions From Stress-Induced Dauphiné Twinning in Quartz},
   volume = {4},
   year = {1979}
}

@article{Chalmers1932,
   author = {Bruce Chalmers},
   journal = {Nature},
   pages = {650-651},
   title = {The cry of tin},
   volume = {129},
   year = {1932}
}

@book{Weertman1964,
   author = {Johannes Weertman and Julia Weertman},
   city = {Cambridge},
   pages = {1-260},
   publisher = {Cambridge University Press},
   title = {Elementary Dislocation Theory},
   year = {1964}
}

@article{Breithaupt2023,
   author = {Thomas Breithaupt and Richard F Katz and Lars N Hansen and Kathryn M Kumamoto},
   doi = {10.1073/pnas},
   issue = {8},
   journal = {Proceedings of the National Academy of Sciences},
   pages = {1-7},
   title = {Dislocation theory of steady and transient creep of crystalline solids: Predictions for olivine},
   volume = {120},
   url = {https://doi.org/10.1073/pnas.2203448120},
   year = {2023}
}

@article{Tommasi2009,
   author = {Andréa Tommasi and Mickael Knoll and Alain Vauchez and Javier W. Signorelli and Catherine Thoraval and Roland Logé},
   doi = {10.1038/ngeo528},
   issn = {17520894},
   issue = {6},
   journal = {Nature Geoscience},
   month = {6},
   pages = {423-427},
   title = {Structural reactivation in plate tectonics controlled by olivine crystal anisotropy},
   volume = {2},
   year = {2009}
}

@article{Tommasi2001,
   author = {Andrëa Tommasi and Alain Vauchez},
   journal = {Earth and Planetary Science Letters},
   pages = {199-210},
   title = {Continental rifting parallel to ancient collisional belts: an e¡ect of the mechanical anisotropy of the lithospheric mantle},
   volume = {185},
   url = {www.elsevier.com},
   year = {2001}
}

@article{Mainprice2005,
   author = {David Mainprice and Andréa Tommasi and Hélène Couvy and Patrick Cordier and Daniel J. Frost},
   doi = {10.1038/nature03266},
   issn = {00280836},
   issue = {7027},
   journal = {Nature},
   month = {2},
   pages = {731-733},
   title = {Pressure sensitivity of olivine slip systems and seismic anisotropy of Earth's upper mantle},
   volume = {433},
   year = {2005}
}

@article{Muto2019,
   author = {J Muto and J D P Moore and S Barbot and T Iinuma and Y Ohta and H Iwamori},
   journal = {Science Advances},
   pages = {1-8},
   title = {Coupled afterslip and transient mantle flow after the 2011 Tohoku earthquake},
   volume = {5},
   url = {https://www.science.org},
   year = {2019}
}

@article{Masuti2023,
   author = {Sagar Masuti and Jun Muto and Erik Rybacki},
   journal = {Journal of Geophysical Research: Solid Earth},
   pages = {1-16},
   title = {Transient Creep of Quartz and Granulite at High Temperature Under Wet Conditions},
   volume = {128},
   year = {2023}
}

@article{Masuti2016,
   author = {Sagar Masuti and Sylvain D. Barbot and Shun Ichiro Karato and Lujia Feng and Paramesh Banerjee},
   doi = {10.1038/nature19783},
   issn = {14764687},
   issue = {7625},
   journal = {Nature},
   pages = {371-377},
   publisher = {Nature Publishing Group},
   title = {Upper-mantle water stratification inferred from observations of the 2012 Indian Ocean earthquake},
   volume = {538},
   year = {2016}
}

@article{Hansen2021,
   author = {Lars N. Hansen and David Wallis and Thomas Breithaupt and Christopher A. Thom and Imogen Kempton},
   doi = {10.1029/2020JB021325},
   issn = {21699356},
   issue = {5},
   journal = {Journal of Geophysical Research: Solid Earth},
   month = {5},
   publisher = {Blackwell Publishing Ltd},
   title = {Dislocation Creep of Olivine: Backstress Evolution Controls Transient Creep at High Temperatures},
   volume = {126},
   year = {2021}
}

@article{Miranda2016,
   author = {Elena A. Miranda and Greg Hirth and Barbara E. John},
   doi = {10.1016/j.jsg.2016.09.002},
   issn = {01918141},
   journal = {Journal of Structural Geology},
   month = {11},
   pages = {30-45},
   publisher = {Elsevier Ltd},
   title = {Microstructural evidence for the transition from dislocation creep to dislocation-accommodated grain boundary sliding in naturally deformed plagioclase},
   volume = {92},
   year = {2016}
}

@article{Montardi1987,
   author = {Y. Montardi and D. Mainprice},
   doi = {10.3406/bulmi.1987.8022},
   issn = {0180-9210},
   issue = {1},
   journal = {Bulletin de Mineralogie},
   pages = {1-14},
   title = {A transmission electron microscopic study of natural plastic deformation of calcic plagioclases (An68-70).},
   volume = {110},
   year = {1987}
}

@article{Ji1990,
   author = {Shaocheng Ji and David Mainprice},
   journal = {Journal of Geology},
   pages = {65-79},
   title = {Recrystallization and fabric development in plagioclase},
   volume = {98},
   url = {http://www.journals.uchicago.edu/t-and-c},
   year = {1990}
}

@article{Jennings2011,
   author = {Andrew T. Jennings and Ju Li and Julia R. Greer},
   doi = {10.1016/j.actamat.2011.05.038},
   issn = {13596454},
   issue = {14},
   journal = {Acta Materialia},
   month = {8},
   pages = {5627-5637},
   title = {Emergence of strain-rate sensitivity in Cu nanopillars: Transition from dislocation multiplication to dislocation nucleation},
   volume = {59},
   year = {2011}
}

@article{Idrissi2016,
   author = {Hosni Idrissi and Caroline Bollinger and Francesca Boioli and Dominique Schryvers and Patrick Cordier},
   doi = {10.1126/sciadv.1501671},
   issn = {23752548},
   issue = {3},
   journal = {Science Advances},
   month = {3},
   pmid = {26998522},
   publisher = {American Association for the Advancement of Science},
   title = {Low-temperature plasticity of olivine revisited with in situ TEM nanomechanical testing},
   volume = {2},
   year = {2016}
}

@article{Korte2011,
   author = {S. Korte and W. J. Clegg},
   doi = {10.1080/14786435.2010.505179},
   issn = {14786435},
   issue = {7-9},
   journal = {Philosophical Magazine},
   title = {Discussion of the dependence of the effect of size on the yield stress in hard materials studied by microcompression of MgO},
   volume = {91},
   year = {2011}
}

@article{Dou2009,
   author = {R. Dou and B. Derby},
   doi = {10.1016/j.scriptamat.2009.05.012},
   issn = {13596462},
   issue = {5},
   journal = {Scripta Materialia},
   month = {9},
   pages = {524-527},
   title = {A universal scaling law for the strength of metal micropillars and nanowires},
   volume = {61},
   year = {2009}
}

@article{Namur2017,
   author = {Olivier Namur and Bernard Charlier},
   doi = {10.1038/ngeo2860},
   issn = {17520908},
   issue = {1},
   journal = {Nature Geoscience},
   month = {1},
   pages = {9-13},
   publisher = {Nature Publishing Group},
   title = {Silicate mineralogy at the surface of Mercury},
   volume = {10},
   year = {2017}
}

@article{Wieczorek2001,
   author = {M. A. Wieczorek and M. T. Zuber},
   doi = {10.1029/2001GL012918},
   issn = {00948276},
   issue = {21},
   journal = {Geophysical Research Letters},
   month = {11},
   pages = {4023-4026},
   title = {The composition and origin of the lunar crust: Constraints from central peaks and crustal thickness modeling},
   volume = {28},
   year = {2001}
}

@article{McSween2009,
   author = {Harry Y McSween and G Jeffrey Taylor and Michael B Wyatt},
   journal = {Science},
   month = {5},
   pages = {736-739},
   title = {Elemental Composition of the Martian Crust},
   volume = {324},
   year = {2009}
}

@article{Collinet2025,
   author = {Max Collinet and Julia Maia and Ana Catalina Plesa and Stephan Klemme and Mark Wieczorek},
   doi = {10.1029/2025JE009187},
   issn = {21699100},
   issue = {11},
   journal = {Journal of Geophysical Research: Planets},
   month = {11},
   publisher = {John Wiley and Sons Inc},
   title = {Felsic Magmatism on Venus Generated by Crustal Recycling and Melting},
   volume = {130},
   year = {2025}
}

@article{Wallis2026,
   author = {David Wallis and Kathryn M Kumamoto and Thomas Breithaupt},
   doi = {10.1038/s41561-026-01920-7},
   issn = {1752-0908},
   issue = {March},
   journal = {Nature Geoscience},
   publisher = {Springer US},
   title = {Mild-to-wild plasticity of Earth's upper mantle},
   volume = {19},
   url = {http://dx.doi.org/10.1038/s41561-026-01920-7},
   year = {2026}
}

@article{Rizzardi2022,
   author = {Q. Rizzardi and C. McElfresh and G. Sparks and D. D. Stauffer and J. Marian and R. Maaß},
   doi = {10.1038/s41467-022-28477-4},
   issn = {20411723},
   issue = {1},
   journal = {Nature Communications},
   pages = {1-9},
   pmid = {35197454},
   publisher = {Springer US},
   title = {Mild-to-wild plastic transition is governed by athermal screw dislocation slip in bcc Nb},
   volume = {13},
   year = {2022}
}

@article{Nespolo2025,
   author = {Massimo Nespolo and Bernd Souvignier},
   doi = {10.1007/s00710-025-00950-0},
   issn = {14381168},
   journal = {Mineralogy and Petrology},
   pages = {161-185},
   title = {Structural analysis of twins in feldspars. II. Albite-pericline twinning and the metric pseudo-symmetry of feldspar lattice},
   year = {2025}
}

@article{Qu2025,
   author = {Tongzhang Qu and Nicolas Brantut and David Wallis and Christopher Harbord},
   doi = {10.1029/2024JB030914},
   issn = {21699356},
   issue = {8},
   journal = {Journal of Geophysical Research: Solid Earth},
   month = {8},
   publisher = {John Wiley and Sons Inc},
   title = {Microstructural Evolution of Carrara Marble During Semi-Brittle Deformation},
   volume = {130},
   year = {2025}
}

@article{Greer2005,
   author = {Julia R. Greer and Warren C. Oliver and William D. Nix},
   doi = {10.1016/j.actamat.2004.12.031},
   issn = {13596454},
   issue = {6},
   journal = {Acta Materialia},
   month = {4},
   pages = {1821-1830},
   title = {Size dependence of mechanical properties of gold at the micron scale in the absence of strain gradients},
   volume = {53},
   year = {2005}
}

@article{Prasad2014,
   author = {K. Eswar Prasad and K. Rajesh and U. Ramamurty},
   doi = {10.1016/j.actamat.2013.10.073},
   issn = {13596454},
   journal = {Acta Materialia},
   month = {2},
   pages = {316-325},
   publisher = {Elsevier Ltd},
   title = {Micropillar and macropillar compression responses of magnesium single crystals oriented for single slip or extension twinning},
   volume = {65},
   year = {2014}
}

@article{DeBresser1993,
   author = {J. H. P. DeBresser and C. J. Spiers},
   doi = {10.1029/92jb02044},
   issn = {0148-0227},
   issue = {B4},
   journal = {Journal of Geophysical Research: Solid Earth},
   pages = {6397-6409},
   title = {Slip systems in calcite single crystals deformed at 300–800°C},
   volume = {98},
   year = {1993}
}

@article{Incel2025,
   author = {Sarah Incel and Markus Ohl and Frans Aben and Oliver Plümper and Nicolas Brantut},
   journal = {Nature Communications},
   pages = {1-13},
   publisher = {Springer US},
   title = {Nanostructures as indicator for deformation dynamics},
   volume = {16},
   year = {2025}
}

@article{Wallis2025,
   author = {David Wallis and Thomas Breithaupt and Taco Broerse},
   doi = {10.1029/2024JB030606},
   issn = {21699356},
   issue = {6},
   journal = {Journal of Geophysical Research: Solid Earth},
   month = {6},
   publisher = {John Wiley and Sons Inc},
   title = {Transient and Steady-State Dislocation Creep of Olivine Controlled by Dislocation Interactions at the Isostress Endmember},
   volume = {130},
   year = {2025}
}

@article{Brown1986,
   author = {William L Brown and Jean Macaudière},
   journal = {Contrib Mineral Petrol},
   pages = {44-56},
   title = {Mechanical twinning of plagioclase in a deformed meta-anorthosite – the production of M-twinning},
   volume = {92},
   year = {1986}
}

@article{Brown2016,
   author = {J. Michael Brown and Ross J. Angel and Nancy L. Ross},
   doi = {10.1002/2015JB012736},
   issn = {21699356},
   issue = {2},
   journal = {Journal of Geophysical Research: Solid Earth},
   month = {2},
   pages = {663-675},
   publisher = {Blackwell Publishing Ltd},
   title = {Elasticity of plagioclase feldspars},
   volume = {121},
   year = {2016}
}

@article{Baisset2024,
   author = {Marie Baïsset and Loïc Labrousse and Alexandre Schubnel and Julien Gasc and Keevin Béneut and Maxime Guillaumet},
   doi = {10.1016/j.jsg.2023.105010},
   issn = {01918141},
   journal = {Journal of Structural Geology},
   month = {1},
   publisher = {Elsevier Ltd},
   title = {Rheology of hydrated plagioclase at lower crustal conditions: Cataclasis, creep and transformational plasticity},
   volume = {178},
   year = {2024}
}

@article{Turner1954,
   author = {F. J. Turner and D.T. Griggs and H. Heard},
   issue = {September},
   journal = {Bulletin of the Geological Society of America},
   pages = {883-934},
   pmid = {38252772},
   title = {Experimental Deformation of Calcite Crystals},
   volume = {65},
   year = {1954}
}

@article{Hansen2019,
   author = {Lars N. Hansen and Kathryn M. Kumamoto and Christopher A. Thom and David Wallis and William B. Durham and David L. Goldsby and Thomas Breithaupt and Cameron D. Meyers and David L. Kohlstedt},
   doi = {10.1029/2018JB016736},
   issn = {21699356},
   issue = {6},
   journal = {Journal of Geophysical Research: Solid Earth},
   pages = {5427-5449},
   title = {Low-Temperature Plasticity in Olivine: Grain Size, Strain Hardening, and the Strength of the Lithosphere},
   volume = {124},
   year = {2019}
}

@article{Uchic2004,
   author = {Michael D. Uchic and Dennis M. Dimiduk and Jeffrey N. Florando and William D. Nix},
   doi = {10.1126/science.1098993},
   issn = {00368075},
   issue = {5686},
   journal = {Science},
   pages = {986-989},
   pmid = {15310897},
   title = {Sample dimensions influence strength and crystal plasticity},
   volume = {305},
   year = {2004}
}

@book{Karato2008,
   author = {Shun-Ichiro Karato},
   city = {Cambridge},
   edition = {First Edit},
   isbn = {978-0-521-84404-8},
   pages = {-463},
   publisher = {Cambridge University Press},
   title = {Deformation of Earth Materials},
   year = {2008}
}

@article{Xie2019,
   author = {Dongyue Xie and Greg Hirth and J. P. Hirth and Jian Wang},
   doi = {10.1007/s00269-019-01055-9},
   issn = {14322021},
   issue = {10},
   journal = {Physics and Chemistry of Minerals},
   month = {11},
   pages = {959-975},
   publisher = {Springer Verlag},
   title = {Defects in deformation twins in plagioclase},
   volume = {46},
   year = {2019}
}

@article{Yund1990,
   author = {R. A. Yund and M. L. Blanpied and T. E. Tullis and J. D. Weeks},
   doi = {10.1029/jb095ib10p15589},
   issn = {01480227},
   issue = {B10},
   journal = {Journal of Geophysical Research},
   pages = {15589-15602},
   title = {Amorphous material in high strain experimental fault gouges},
   volume = {95},
   year = {1990}
}

@article{Sly2020,
   author = {Michael K. Sly and Arashdeep S. Thind and Rohan Mishra and Katharine M. Flores and Philip Skemer},
   doi = {10.1093/gji/ggz577},
   issn = {1365246X},
   issue = {1},
   journal = {Geophysical Journal International},
   month = {4},
   pages = {129-141},
   publisher = {Oxford University Press},
   title = {Low-temperature rheology of calcite},
   volume = {221},
   year = {2020}
}

@article{Rowe1990,
   author = {K. J. Rowe and E. H. Rutter},
   journal = {Journal of Structural Geology},
   pages = {17},
   title = {Palaeostress estimation using calcite twinning: experimental calibration and application to nature},
   volume = {12},
   year = {1990}
}

@article{Pec2021,
   author = {Matej Pec and Saleh Al Nasser},
   doi = {10.1029/2020JB021262},
   issn = {21699356},
   issue = {5},
   journal = {Journal of Geophysical Research: Solid Earth},
   month = {5},
   pages = {1-26},
   publisher = {Blackwell Publishing Ltd},
   title = {Formation of Nanocrystalline and Amorphous Materials Causes Parallel Brittle-Viscous Flow of Crustal Rocks: Experiments on Quartz-Feldspar Aggregates},
   volume = {126},
   year = {2021}
}

@article{Olsen1984a,
   author = {Terkel S Olsen and David L Kohlstedt},
   journal = {Physics and Chemistry of Minerals},
   pages = {153-160},
   title = {Analysis of Dislocations in Some Naturally Deformed Plagioclase Feldspars},
   volume = {11},
   year = {1984}
}

@article{Lacombe2021a,
   author = {Olivier Lacombe and Camille Parlangeau and Nicolas E. Beaudoin and Khalid Amrouch},
   doi = {10.3390/geosciences11110445},
   issn = {20763263},
   journal = {Geosciences},
   publisher = {MDPI},
   title = {Calcite twin formation, measurement and use as stress–strain indicators: A review of progress over the last decade},
   volume = {11},
   year = {2021}
}

@article{Lacombe2001,
   author = {Olivier Lacombe},
   doi = {10.1029/2001TC900019},
   issn = {02787407},
   issue = {6},
   journal = {Tectonics},
   month = {12},
   pages = {834-849},
   publisher = {Blackwell Publishing Ltd},
   title = {Paleostress magnitudes associated with development of mountain belts: Insights from tectonic analyses of calcite twins in the Taiwan foothills},
   volume = {20},
   year = {2001}
}

@article{Korte-Kerzel2017,
   author = {Sandra Korte-Kerzel},
   doi = {10.1557/mrc.2017.15},
   issn = {21596867},
   issue = {2},
   journal = {MRS Communications},
   month = {6},
   pages = {109-120},
   publisher = {Cambridge University Press},
   title = {Microcompression of brittle and anisotropic crystals: Recent advances and current challenges in studying plasticity in hard materials},
   volume = {7},
   year = {2017}
}

@article{Lacombe2007a,
   author = {Olivier Lacombe},
   doi = {10.1016/j.jsg.2006.08.009},
   issn = {01918141},
   issue = {1},
   journal = {Journal of Structural Geology},
   month = {1},
   pages = {86-99},
   title = {Comparison of paleostress magnitudes from calcite twins with contemporary stress magnitudes and frictional sliding criteria in the continental crust: Mechanical implications},
   volume = {29},
   year = {2007}
}

@article{Korte2009,
   author = {S. Korte and W. J. Clegg},
   doi = {10.1016/j.scriptamat.2009.01.029},
   issn = {13596462},
   issue = {9},
   journal = {Scripta Materialia},
   month = {5},
   pages = {807-810},
   title = {Micropillar compression of ceramics at elevated temperatures},
   volume = {60},
   year = {2009}
}

@article{Mehl2008,
   author = {Luc Mehl and Greg Hirth},
   doi = {10.1029/2007JB005075},
   issn = {21699356},
   issue = {5},
   journal = {Journal of Geophysical Research: Solid Earth},
   month = {5},
   publisher = {Blackwell Publishing Ltd},
   title = {Plagioclase preferred orientation in layered mylonites: Evaluation of flow laws for the lower crust},
   volume = {113},
   year = {2008}
}

@article{Kranjc2020,
   author = {Kelly Kranjc and Arashdeep S. Thind and Albina Y. Borisevich and Rohan Mishra and Katharine M. Flores and Philip Skemer},
   doi = {10.1029/2019JB019242},
   issn = {21699356},
   issue = {5},
   journal = {Journal of Geophysical Research: Solid Earth},
   month = {5},
   publisher = {Blackwell Publishing Ltd},
   title = {Amorphization and Plasticity of Olivine During Low-Temperature Micropillar Deformation Experiments},
   volume = {125},
   year = {2020}
}

@article{DeBresser1997,
   author = {J H P DeBresser and C J Spiers},
   journal = {Tectonophysics},
   pages = {1-23},
   title = {Strength characteristics of the r, f, and c slip systems in calcite},
   volume = {272},
   year = {1997}
}

@article{Incel2024,
   author = {Sarah Incel and Lisa Katharina Mohrbach and Jörg Renner},
   doi = {10.1029/2023GC011275},
   journal = {Geochemistry, Geophysics, Geosystems},
   pages = {1-15},
   title = {How Strong / Weak Is Epidote Relative to Plagioclase ?},
   volume = {25},
   year = {2024}
}

@article{Rutter2022,
   author = {Ernest Rutter and David Wallis and Kamil Kosiorek},
   doi = {10.3390/GEOSCIENCES12060222},
   issn = {20763263},
   issue = {6},
   journal = {Geosciences (Switzerland)},
   month = {6},
   publisher = {MDPI},
   title = {Application of Electron Backscatter Diffraction to Calcite-Twinning Paleopiezometry},
   volume = {12},
   year = {2022}
}

@article{Paterson1958,
   author = {M. S. Paterson},
   journal = {Bulletin of the Geological Society of America},
   title = {Experimental deformation and faulting in Wombeyan marble},
   volume = {69},
   year = {1958}
}

@article{Behr2014a,
   author = {Whitney M Behr and John P Platt},
   doi = {10.1002/2014GL061349.Received},
   journal = {Geophysical Research Letters},
   pages = {8067-8075},
   title = {Brittle faults are weak, yet the ductile middle crust is strong : Implications for lithospheric mechanics},
   volume = {41},
   year = {2014}
}

@article{Ronov1969,
   author = {A. B. Ronov and A. A. Yaroshevsky},
   journal = {Geophysical Monograph Series},
   pages = {37-57},
   title = {Chemical Composition of the Earth's Crust},
   volume = {13},
   year = {1969}
}

@article{Christian1995,
   author = {J. W. Christian and S. Mahajan},
   doi = {10.1016/0079-6425(94)00007-7},
   journal = {Progress in Materials Science},
   pages = {1-157},
   publisher = {Elsevier Ltd},
   title = {Deformation twinning},
   volume = {39},
   year = {1995}
}

@article{Olsen1985,
   author = {T. S. Olsen and David L. Kohlstedt},
   journal = {Tectonophysics},
   pages = {107-131},
   title = {Natural deformation and recrystallization of some intermediate plagioclase feldspars},
   volume = {111},
   year = {1985}
}

@article{Incel2019,
   author = {Sarah Incel and Loïc Labrousse and Nadège Hilairet and Timm John and Julien Gasc and Feng Shi and Yanbin Wang and Torgeir B Andersen and François Renard and Bjørn Jamtveit and Alexandre Schubnel},
   doi = {https://doi.org/10.1130/G45527.1},
   issue = {3},
   journal = {Geology},
   pages = {235-238},
   title = {Reaction-induced embrittlement of the lower continental crust},
   volume = {47},
   year = {2019}
}

@article{Stunitz2003,
   author = {H Stünitz and J D Fitz Gerald and J Tullis},
   doi = {10.1016/S0040-1951(03)00241-5},
   journal = {Tectonophysics},
   pages = {215-233},
   title = {Dislocation generation , slip systems , and dynamic recrystallization in experimentally deformed plagioclase single crystals},
   volume = {372},
   year = {2003}
}

@article{Rybacki2000,
   author = {E Rybacki and G Dresen},
   doi = {10.1029/2000JB900223},
   journal = {Journal of Geophysical Research},
   pages = {26,017-26,036},
   title = {Dislocation and diffusion creep of synthetic anorthite aggregates},
   volume = {105},
   year = {2000}
}

@article{Tullis1992,
   author = {J. Tullis and R. Yund},
   doi = {10.1016/S0074-6142(08)62816-8},
   issue = {C},
   journal = {International Geophysics},
   pages = {89-117},
   title = {The Brittle-Ductile Transition in Feldspar Aggregates: An Experimental Study},
   volume = {51},
   year = {1992}
}

@article{Marshall1977,
   author = {Db Marshall and Ac McLaren},
   doi = {10.1007/BF00308845},
   issue = {August 1976},
   journal = {Physics and chemistry of minerals},
   pages = {351-370},
   title = {Deformation mechanisms in experimentally deformed plagioclase feldspars},
   volume = {370},
   year = {1977}
}

@article{Borg1969,
   author = {I. Y. Borg and H. C. Heard},
   doi = {10.1007/BF00375174},
   issn = {00107999},
   issue = {2},
   journal = {Contributions to Mineralogy and Petrology},
   pages = {128-135},
   title = {Mechanical twinning and slip in experimentally deformed plagioclases},
   volume = {23},
   year = {1969}
}

@article{Pec2012a,
   author = {Matej Pec and Holger Stünitz and Renée Heilbronner and Martyn Drury and Christian de Capitani},
   doi = {10.1016/j.epsl.2012.09.004},
   isbn = {0012-821X},
   issn = {0012821X},
   journal = {Earth and Planetary Science Letters},
   pages = {299-310},
   title = {Origin of pseudotachylites in slow creep experiments},
   volume = {355-356},
   year = {2012}
}

@article{Pabst2015,
   author = {Willi Pabst and Eva Gregorová and Elisa Rambaldi and Maria Chiara Bignozzi},
   issn = {18045847},
   issue = {4},
   journal = {Ceramics - Silikaty},
   pages = {326-330},
   title = {Effective elastic constants of plagioclase feldspar aggregates in dependence of the anorthite content - A concise review},
   volume = {50},
   year = {2015}
}

\pagebreak

\subsection*{Acknowledgement}
This study greatly benefitted from numerous discussions with Nicolas Brantut (NB), and received funding to S.K.-K. and N.B. from the European Research Council (ERC) under the European Union’s Horizon research and innovation programme (101168203/"TailorPlast"; 101088963/“RockDeath”). S.I. is very grateful to Jörg Renner, Evangelos Korkolis, and Katharina (Tinka) Marquardt, for fruitful discussions, and thanks the team at Alemnis AG, especially Renato Pero and Miguel Herrero, for their support and input. S.I. further thanks Hans-Peter Schertl for providing her with the anorthite crystal used for this study. S.I. acknowledges UKRI for this research received funding in form of a NERC Independent Research Fellowship (Grant Agreement NE/ W00805X/1).

\end{document}